# Projection-Based Reduced Order Model for Simulations of Nonlinear Flows with Multiple Moving Objects

My Ha DAO

Institute of High Performance Computing, Agency for Science, Technology and Research,
1 Fusionopolis Way, #16-16 Connexis, Singapore 138632.
daomh@ihpc.a-star.edu.sg

ABSTRACT

This paper presents a reduced order approach for transient modeling of multiple moving objects in nonlinear crossflows. The Proper Orthogonal Decomposition method and the Galerkin projection are used to construct a reduced version of the nonlinear Navier-Stokes equations. The Galerkin projection implemented in OpenFOAM platform allows accurate impositions of arbitrary time-dependent boundary conditions at the moving boundaries. A modelling technique based on moving domain and immersed boundary techniques is proposed to overcome the challenge of handling moving boundaries due to movements of the multiple objects. The model is demonstrated capable to capture the complex flow fields past one and two oscillating cylinders and the forces acting on the cylinders. Simulation time could be reduced by more than 1000% for a small case on a fine mesh as compared to an existing method and could be more for large cases. In general, the simulation time of the reduced model is of order of seconds as compared to hours of the full order Computational Fluid Dynamics models.

## 1 Introduction

The Reduced Order Modeling (ROM) approach based on the Proper Orthogonal Decomposition (POD) method has been extensively adopted for various engineering applications. The fundamental behind the POD method is the autocorrelation of a set of numerical results of state variables such as velocity and pressure, called snapshots, on a fixed computational mesh [1-5] and solving an eigen-problem of the autocorrelation matrix. The POD-based ROM is an established methodology for both steady-state and transient simulations in fluid dynamics, thermodynamics, structural mechanics, etc [6-8]. The method has also been applied to various large scale engineering problems such as a reservoir simulation [9], ocean engineering [10]. Various POD methods have also been developed to handle different challenging aspects in ROM. These include an adaptive POD method to incorporate unseen solutions into the POD basis vectors [11], a Galerkin-projection to construct ROM through a projections of governing equations on the POD basis vectors [12] or a Discrete Empirical Interpolation method (DEIM) to handle high-order nonlinear terms in governing equations [13]. Approaches for constructing hybrid ROM and Machine Learning are also explored [14]. In all these developements, the computational domain is unchanged. Very few work has addressed the issue of multiple moving objects in a crossflow which can be encountered in many engineering applications where both flow field and loading are of importance. Some examples are the wind shielding effect of a Floating Liquefied Natural Gas (FLNG) vessel to a LNG carrier in operation, the vibration of a riser in a wake of other risers. The numerical simulation of these problems often requires special treatments to handle moving boundaries such as moving computational meshes. Even if the mesh topology (i.e. the order and connectivity of mesh points) is unchanged, locations of mesh points may change significantly in every snapshot. That affects the validity of the autocorrelation and hence the accuracy of the ROM models. Here we focus on the ROM development targeting to simulate crossflows with moving objects. For comprehensive reviews of POD and ROM methodologies and applications readers are recommended to the work in [15-17].

There have been several developments to handle the moving mesh for a single object in ROM. The simplest technique introduced in [18] is to build multiple ROMs for different sets of mesh deformation and uses a mesh metric to determine when to switch between ROMs. This method is suitable for arbitrary mesh deformations, but an invariant mesh is still need for every set of mesh deformation to construct the ROM, and multiple ROM models are required. For problems with small mesh deformation, actuation mode method could be used [19-21]. The computational mesh is basically unchanged, while actuation modes as functions of the displacement of the object's boundaries are introduced into the POD modes to model the effect of object's movement to the flow field in the vicinity. This method is only valid for small displacement problems and particularly suitable for modelling small vibration of the object's surface. In [22], a domain with moving boundaries was mapped into a stationary domain prior to the decomposition. The mapping comprises a transfinite interpolation and a volume adjustment algorithm. The ROM solutions are then mapped back to the original domain for final predictions. Hence, a prerequisite condition that the forward and inverse mapping must be identified is required for this method. In [23], the governing equations used in the construction of ROM are rewritten with modified primitive variables which relates to the velocity of the object in a moving frame (or reference mesh). The reference mesh is fixed with respect to the object and is deformable. The introduction of the primitive variables leads to the apparition of the Coriolis and centrifugal forces in the reference mesh that are considered as source terms. This method is similar to the dynamics mesh technique in the physics-based numerical methods [24] in which the velocity of every mesh points is solved in parallel with the flow field.

The above methods have been demonstrated working well for certain problems involving a single object. An extension of these method to complex scenarios such as multiple moving objects can be impractical because defining a simple reference mesh fixed to the objects or a deterministic mesh transformation is not always possible. There are existing surrogate methods to predict complex mesh velocity based on the deformations of computational domains which can be coupled with the ROM in [23]. These include Radial Basis Function Interpolation, Inverse Distance Weighting or the POD-based ROM [25-27]. Using these methods will add an additional layer of complexity to the primary ROM model on the primitive variables. In cases of large or complex mesh deformations, these methods may lead to low mesh qualities, as shown in [25-27], and this can be detrimental to the accuracy and stability of the primary ROM model.

Another set of methods that are versatile for handling moving objects are inspired by the Immersed Boundary (IB) method [28] and adopted into ROM by several researchers [29-31]. Prior to the POD decomposition, all solution snapshots are interpolated into a fictitious stationary domain that includes both the fluid and the object. The effects of the object's motions on the fluid are modelled by additional forcing terms in the ROM equations. The computation of the force terms, though in slightly different ways, are all similar to the IB method and could be complex. These methods are flexible and capable to handle large and complex motions of the objects (including objects collision, though this is not considered in this work). Similar to the original IB method, the accuracy of the resultant ROMs is highly dependent on the computations of the force terms. Beside the concern of model accuracy, in the ROM context, a complex computation of the force terms could significantly increase computational cost and hence defeats the benefits of using ROM.

Knowing the shortcomings of the reference moving frame and the IB-like methods in ROM simulations, this work attempts to extend the two methods to into a general framework for modeling multiple moving objects. An implementation of the more accurate reference moving mesh where possible and a simplified IB method would allow a fast simulation of multiple moving objects while focusing the model accuracy at the object of interest. Although, the reference mesh and IB-like methods are adopted, several modifications to the methods to improve their efficiency are implemented.

Firstly, the moving frame method similar to the one used in [23] where the reference mesh is fixed with respect to the rigid object is considered. Instead of using a deformable mesh and primitive variables which leads to additional source terms (hence additional on-line computational cost), the mesh is non-deformable and the off-line data are interpolated into the reference mesh before applying POD. This approach is, however, different from the domain mapping method used in [22] as no mesh transformation is required. The effects of moving mesh are applied directly at the boundaries of the reference mesh as velocity boundary conditions. Secondly, the IB-like method [30] is reviewed and modified to improve its efficiency based on analytical and numerical analyses. Next, these algorithms will be built on top of the PBROM framework that are based on the OpenFOAM numerical platform. This PBROM framework makes use of the OpenFOAM's well developed and validated numerical discretization schemes and the ability to handle arbitrary, time-dependent boundary conditions (both Dirichlet and Neumann) at all domain boundaries [32], which are critical for the development of the reference mesh technique. And finally, these algorithms are tested on various test cases for the capability of capturing transient flow fields and, more importantly, the forces acting on the moving objects. For the demonstration of the algorithms, a crossflow of Re = 100 is considered and the objects' movements are prescribed. Although the test cases are relatively simplified as compared to realistic problems, this work is probably among the very few attempts to extent ROM's capability to model multiple moving objects as well as to capture the forces acting on the objects. The development of PBROM for simulations of high Re flow is beyond the focus of this paper and is currently conducted in a separate study by the author.

The paper is organized as follows. Brief descriptions of the PBROM and its implementation in OpenFOAM [32] are presented in Section 2.1 – 2.3. Implementations of moving frame and IB-like methods in the PBROM are presented Section 2.4. The PBROM is demonstrated on predictions of flow fields and forces on a single and two moving cylinders in Section 3.

## 2 Projection-based ROM for incompressible flows

### 2.1 Full model of incompressible flows in OpenFOAM

Incompressible flows past moving objects are governed by the Navier-Stokes (NS) equations which, in Cartesian coordinates $\mathbf{x} = (x, y, z)$ and Arbitrary Eulerian Lagrangian (ALE) framework, have the form of

$$\mathbf{u}_t = -(\mathbf{u} - \mathbf{u}_L) \cdot \nabla \mathbf{u} + \nu \Delta \mathbf{u} - \nabla p/\rho \tag{1}$$

$$\nabla \cdot \mathbf{u} = 0 \tag{2}$$

where the velocity $\mathbf{u} = (u, v, w)$, the pressure $p$ are spatio-temporal functions in a computational domain $\Omega_f$; $\mathbf{u}_L$ is the Lagrangian mesh velocity; $\rho$ and $\nu$ are the density and the kinematic viscosity of the fluid. The governing Partial Differential Equations (PDEs) can be solved by many available Computational Fluid Dynamics (CFD) packages such as Fluent, StarCCM (commercial) or OpenFOAM (open-source).

In this development, the OpenFOAM platform with the "pimpleDyMFoam" solver is employed. The solver uses the Finite Volume Method (FVM) for spatial discretisation, the dynamic mesh techniques for moving objects and the Pressure Implicit with Splitting of Operator (PISO) algorithm to solve the discretised NS equations [33, 34]. Using FVM, the computational domain $\Omega_f$ is divided into a finite number of control volumes $\delta\Omega$, whose centres define the mesh. The gradient ($\nabla$), divergence ($\nabla \cdot$) are computed as

$$\int_{\delta\Omega} \nabla * \omega \, d\Omega = \int_{\delta S} d\mathbf{S} * \omega = \sum_k \mathbf{S}_k * \omega_{\mathbf{S}_k} \tag{3}$$

Here, $\omega$ represents the state variables, e.g. $\omega := u, v, w, p$. The operator $*$ represents inner and outer products for respective differential operators: the divergence $\nabla \cdot \omega$ and the gradient $\nabla \omega$. The nonlinear convection term in the momentum equation is integrated and linearized as

$$\int_{\delta\Omega} \nabla \cdot (\mathbf{u}\omega) \, d\Omega = \int_{\delta S} d\mathbf{S} \cdot (\mathbf{u}\omega)_{d\mathbf{S}} = \sum_k (\mathbf{S}_k \cdot \mathbf{u}_{\mathbf{S}_k})\omega_{\mathbf{S}_k} = \sum_k F_{\mathbf{S}_k} \omega_{\mathbf{S}_k} \tag{4}$$

where the face flux, $F_{\mathbf{S}_k}$, defined as the surface normal component of the velocity $F_{\mathbf{S}_k} := \mathbf{S}_k \cdot \mathbf{u}_{\mathbf{S}_k}$, is computed by using the velocity from the previous time step and $\omega_{\mathbf{S}_k}$ is the face value of $\omega$. The Laplacian ($\Delta$) operator involves the divergence of a face gradient $(\nabla\omega)_{d\mathbf{S}}$ and is integrated as

$$\int_{\delta\Omega} \nabla \cdot (\nabla\omega) \, d\Omega = \int_{\delta S} d\mathbf{S} \cdot (\nabla\omega)_{d\mathbf{S}} = \sum_k \mathbf{S}_k \cdot (\nabla\omega)_{\mathbf{S}_k} \tag{5}$$

The face gradient on face $k$ of the control volume $i$ is calculated as a linear function of the values at the centres of cell $i$ and surrounding cells $j$ and the length vector $\mathbf{d}$ linking the two cell centres,

$$\mathbf{S}_k \cdot (\nabla \omega)_{\mathbf{S}_k} = |\mathbf{S}_k| \frac{\omega_j - \omega_i}{|\mathbf{d}|} \tag{6}$$

While the solution is defined at centres of control volumes, the boundary values are defined at centres of boundary faces. For Dirichlet boundary conditions, prescribed face values are simply substituted into the discretisation where required, e.g. in Eq. (4) and (6). For Neumann boundary conditions, prescribed gradients are also substituted into the discretisation where required, e.g. in Eq. (5). Where the discretisation requires a boundary face value, it is interpolated from cell centre value as

$$\omega_{\mathbf{S}_k} = \omega_i + \frac{\mathbf{S}_k \cdot (\nabla \omega)_{\mathbf{S}_k}}{|\mathbf{S}_k|} |\mathbf{d}| \tag{7}$$

## 2.2 Proper Orthogonal Decomposition (POD)

The POD method was introduced in [35] and has also been referred to as Principal Component Analysis [36] and Karhunen–Loève decomposition [37]. Given an ensemble of $m$ vectors of the variables (or "snapshots"), $\mathbf{X} := \{\omega_i\}_{i=1}^m \in \mathbb{R}^N$, such as the velocity component or the pressure obtained from CFD simulations, each snapshot can be represented in a subspace $S$ in the form

$$\omega = \sum_{i=1}^m a_i \phi_i \tag{8}$$

Here, $\{a_i\}_{i=1}^m \in \mathbb{R}$ are the modal coefficients and $\{\phi_i\}_{i=1}^m \in \mathbb{R}^N$ are the basis vectors (also called POD modes) spanning the subspace $S$ and $N$ is the number of unknowns of a variable of the CFD model. The basis vectors can be computed using the "method of snapshots" introduced by Sirovich [38] as linear combination of the snapshots,

$$\phi = \sum_{i=1}^m \beta_i \omega_i \tag{9}$$

where the coefficients $\{\beta_i\}_{i=1}^m \in \mathbb{R}$ satisfy the eigen-problem,

$$\mathbf{R}\beta = \lambda \beta \tag{10}$$

where $\mathbf{R} \in \mathbb{R}^{m \times m}$ is a correlation matrix, defined as $\mathbf{R}_{ij} := \frac{1}{m}(\omega_i, \omega_j)$. The eigenvalues $\{\lambda_i\}_{i=1}^m \in \mathbb{R}$ determine the importance of the basis vectors in construction of the snapshots. The set of $r$ basis vectors, $r \ll m$, used to approximate a snapshot are chosen according to the largest corresponding $\lambda$ such that $\sum_{i=1}^r \lambda_i / \sum_{i=1}^m \lambda_i > e_r$, where $e_r$ is called the "relative energy" captured by the chosen $r$ basis vectors. The approximation of a snapshot, $\widetilde{\omega} \approx \omega$, is given by

$$\widetilde{\omega} = \sum_{i=1}^{r} a_i \phi_i \tag{11}$$

In general, the ensemble mean $\omega_0 = \langle \omega \rangle$ is non-zero and often contains the largest "relative energy" of the snapshots. The ensemble mean is therefore often excluded from the POD operation. The approximation (11) could be rewritten as

$$\widetilde{\omega} = \omega_0 + \sum_{i=1}^{r} a_i \phi_i \tag{12}$$

For simplicity, $\widetilde{\omega}$ will be replaced by $\omega$ in the below sections.

## 2.3 Projection based ROM (PBROM) in OpenFOAM

Let $\omega$ be a variable of the PDEs defined in Section 2.1, e.g. $\omega := u, v, w, p$ for the velocity components and pressure, for an incompressible flow in the inner domain $\Omega := \Omega_f \backslash \Gamma$, where $\Gamma$ is the boundary of $\Omega_f$. If a Dirichlet boundary condition is applied on $\Gamma$, i.e. $\omega = g_D$ is pre-determined, then one can re-define the ensemble mean as $\omega_0 := \langle \omega_\Omega \rangle \cup g_D$, allowing the basis vectors to be constructed from an ensemble of flow solutions on the inner domain $\Omega$ that are separated from the boundary values. The approximation (12) will automatically satisfy the boundary conditions on $\Gamma$ for any $\{a_i\}$. That generally does not work with Neumann boundary condition, i.e. $\nabla \omega = g_N$ on $\Gamma$, where $\omega_\Gamma$ and $\omega_\Omega$ are inter-dependent. However, it can be shown that the following statements, (13) and (14), are equivalent when $g_D$ and $\omega_\Omega$ are linearly dependent:

$$\omega_1 := \{\omega_1|_\Omega = \omega_\Omega, \quad \nabla \omega_1|_\Gamma = g_N\} \tag{13}$$

$$\begin{aligned} \omega_{11} &:= \{\omega_{11}|_\Omega = \omega_\Omega, \quad \nabla \omega_{11}|_\Gamma = 0\} \\ \omega_{12} &:= \{\omega_{12}|_\Omega = \omega_\Omega, \quad \nabla \omega_{12}|_\Gamma = 1\} \\ \omega_1 &:= \omega_{11} + (\omega_{12} - \omega_{11})g_N \end{aligned} \tag{14}$$

where $\omega_i, i = 1, 11, 12$, are different fields of $\omega$ on the full computational domain $\Omega_f$. Note that the above equivalence is also applicable for Dirichlet boundary conditions, i.e.

$$\omega_1 := \{\omega_1|_\Omega = \omega_\Omega, \quad \omega_1|_\Gamma = g_D\} \tag{15}$$

$$\begin{aligned} \omega_{11} &:= \{\omega_{11}|_\Omega = \omega_\Omega, \quad \omega_{11}|_\Gamma = 0\} \\ \omega_{12} &:= \{\omega_{12}|_\Omega = \omega_\Omega, \quad \omega_{12}|_\Gamma = 1\} \\ \omega_1 &:= \omega_{11} + (\omega_{12} - \omega_{11})g_D \end{aligned} \tag{16}$$

Given the definitions (13) - (16), the variable $\omega$ could be decomposed in the following form

$$\omega = \omega_0 + \boldsymbol{\Phi}^\omega a^\omega + \mathbf{G}^\omega b^\omega \tag{17}$$

where $\boldsymbol{\Phi}^\omega := \{\phi_i^\omega\}_{i=1}^{r^\omega}$ and $\mathbf{G}^\omega := \{g_i^\omega\}_{i=1}^{q^\omega}$ are the matrices comprising of the basic vectors for the inner domain, $\phi_i^\omega$, and the boundary condition functions for the boundary, $g_i^\omega$, respectively; $a^\omega := \{a_i^\omega\}_{i=1}^{r^\omega}$ and $b^\omega := \{b_i^\omega\}_{i=1}^{q^\omega}$ are the coefficients; $r^\omega$ is the number of basis vectors and $q^\omega$ is the number of boundaries. Comparing (17) with (13) - (16), one can derive $g^\omega = \omega_{12} - \omega_{11}$ and $b^\omega = g_D, g_N$.

To project the NS equations on the solution subspaces, the Poisson equation for pressure is first derived following steps in the PISO algorithm. The time derivative of velocity term in the momentum equation (1) at a time instance $t$ is discretised by a forward Euler scheme, i.e. $\mathbf{u}_t = (\mathbf{u}^{t+dt} - \mathbf{u}^t)/dt$, where $dt$ is the time step. Next, take the divergence of the discretised momentum equation and substitute it into the continuity equation (2). After rearranging and noting that the velocity at time $t + dt$ satisfies the continuity equation (2), i.e. $\nabla \cdot \mathbf{u}^{t+dt} = 0$, one obtains

$$\frac{1}{\rho}\Delta p = \frac{\nabla \cdot \mathbf{u}}{dt} - \nabla \cdot (\mathbf{u} \cdot \nabla \mathbf{u}) + \nu \Delta (\nabla \cdot \mathbf{u}) \tag{18}$$

where $\mathbf{u}$ denotes the velocity at time $t$ or $\mathbf{u}^t$.

The momentum and the pressure Poisson equations with the state variables in their decomposed forms (17) are now projected on the respective solution subspaces spanning $r^\omega$ basis vectors, $S^\omega := \mathbf{\Phi}^\omega$, in an off-line phase to obtain the PBROM which is comprised of a set of Ordinary Differential Equations (ODEs) as

$$a_t^\mathbf{u} = f_\mathbf{u}(a^\mathbf{u}, a^p) + g_\mathbf{u}(b^\mathbf{u}, b^p) \tag{19}$$

$$a^p = f_p(a^\mathbf{u}) + g_p(b^\mathbf{u}, b^p) \tag{20}$$

where $a^\mathbf{u} := [a^u, a^v, a^w]^T$ and $a^p$ are the modal coefficients; $b^\mathbf{u} := [b^u, b^v, b^w]^T$ and $b^p$ are the boundary conditions for velocity and pressure, respectively. The projection employs the numerical schemes in OpenFOAM and the Galerkin projection, which is defined as a cell volume weighted dot-product. The resulted functions $f_\mathbf{u}, g_\mathbf{u}, f_p, g_p$ are provided in detail in the appendix.

Using the PBROM model, i.e. Eqs (19) and (20), the state variables become the modal coefficients for velocities and pressure $a^u, a^v, a^w, a^p$. The number of unknowns is reduced from $4N$ to $R$, where $R = r^u + r^v + r^w + r^p \sim \mathcal{O}(10^2)$. The PB-ROM model can be solved by the predictor-corrector scheme (a simplification of the PISO algorithm in OpenFOAM [34]) as described in ALGORITHM 1.

*Algorithm 1*

Step 1: Equation (19) is solved by a time integration scheme with time step $dt$, the present pressure coefficient $a^p$ and boundary conditions $b^\mathbf{u}, b^p$ to obtain the intermediate velocity coefficient $a^{\mathbf{u}^*}$.

Step 2: Equation (20) is solved to obtain the intermediate pressure coefficient $a^{p^*}$ using the intermediate velocity coefficient $a^{\mathbf{u}^*}$ and the boundary conditions $b^\mathbf{u}, b^p$. The increment of pressure coefficient, $da^p = a^{p^*} - a^p$, is then calculated.

Step 3: Correct the intermediate velocity coefficients, $a^{\mathbf{u}^*} \leftarrow a^{\mathbf{u}^*} + dt f_{\mathbf{u},p}(da^p)$, where $f_{\mathbf{u},p}$ comprises $a^p$-dependent components of $f_\mathbf{u}$.

Step 4: Correction of pressure and velocity coefficients, Step 2 - 3, is repeated until the intermediate velocity converges, i.e. $dt f_{\mathbf{u},p}(da^p) \leq \varepsilon$ or a maximum number of iterations is reached.

## 2.4 Handing moving and deformable objects

Using the OpenFOAM's dynamic mesh techniques [24], positions of cell centres where solution snaphots are collected vary according to the movements of the objects. Hence, a mesh movement problem needs to be handled before applying POD and Galerkin projection. The concept of moving frame modified from those used in [22] and [23] is implemented. A non-deformable reference mesh $\Omega$ is fixed with respect to the rigid object as demonstrated in Figure 1A. The off-line data are interpolated into the reference mesh before applying POD. This method hence does not require mesh transformations nor mesh velocity calculation. The effects of moving mesh are applied directly at the boundaries of the reference mesh as velocity boundary conditions $b^{\mathbf{u}}$ under the functions $g_{\mathbf{u}}$ and $g_p$. In general, both translational and rotational motions of the object can be considered under this framework.

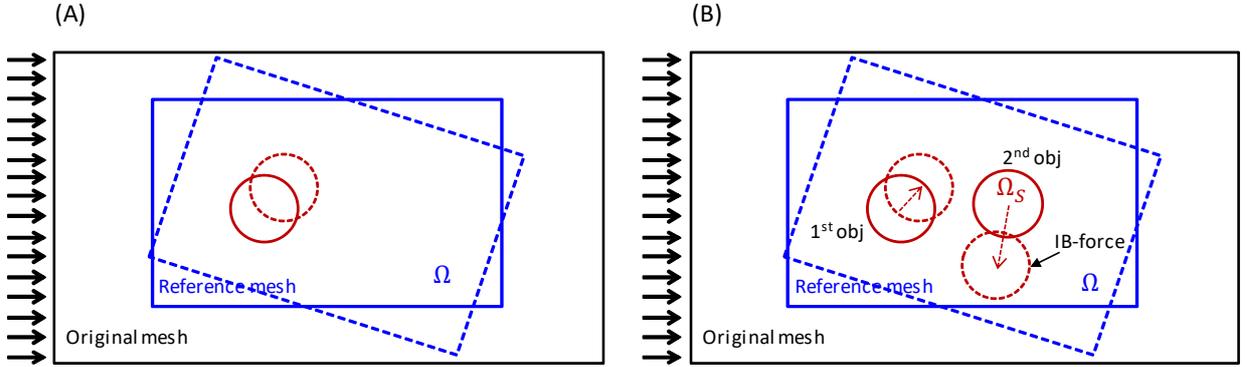

**Figure 1 Schematic demonstration of the moving mesh techniques for ROM: (A) ALE; (B) ALE+IB.**

To model the 2nd moving objects in the PBROM, the IB-like approach is employed. As demonstrated in Figure 1B, a non-deformed moving mesh covering all objects is attached to the 1st object while the 2nd object, $\Omega_S$, is replaced by a fluid domain. The PBROM is now constructed on the extended domain $\Omega \leftarrow \Omega \cup \Omega_S$. The velocity inside $\Omega_S$ is interpolated from the velocity of the object's surface, while pressure is set to the reference pressure. Note that the velocity of the 2nd object must be in relative to the velocity of the moving mesh. Following the idea of the IB method [19], a source term, $\mathbf{F}_{IB}$, representing the effect of the 2nd object on the flow field is added into the original momentum equation,

$$\mathbf{F}_{IB} = \begin{cases} (\mathbf{v}_0 - \mathbf{u})/dt & \text{in } \Omega_S \\ 0 & \text{otherwise} \end{cases} \qquad (21)$$

In the PBROM simulations, the velocity field $\mathbf{u}$ is reconstructed following (17) at the previous time step, and $\mathbf{v}_0$ is the velocity of the 2nd object at the current time step. The $\mathbf{F}_{IB}$ term hence represents the discretized accelerational force of the object acting on the fluid. The IB source term is then projected on the solution subspaces of the velocity and added into the right-hand side of the PBROM momentum equation as

$$a^{\mathbf{u}}_t = f_{\mathbf{u}}(a^{\mathbf{u}}, a^p) + g_{\mathbf{u}}(b^{\mathbf{u}}, b^p) + \mathbf{f}_{IB} \tag{22}$$

$$\mathbf{f}_{IB} = \begin{cases} (\mathbf{v}_0 - \mathbf{u}_0 - \mathbf{\Phi}^{\mathbf{u}} a^{\mathbf{u}} - \mathbf{G}^{\mathbf{u}} b^{\mathbf{u}}, \mathbf{\Phi}^{\mathbf{u}})/dt & \text{in } \Omega_S \\ 0 & \text{otherwise} \end{cases} \tag{23}$$

where $(\cdot,\cdot)$ is the inner product which performs the Galerkin projection of the forcing terms on the solution subspaces. This formula for $\mathbf{F}_{IB}$ is similar to the IB-like formula used in [30] which was said to allow large and arbitrary movement of boundaries. However, compared to the formula in [30], the convection-diffusion term $(\mathbf{u} \cdot \nabla \mathbf{u} - \nu \Delta \mathbf{u})^t$ is dropped out from the calculation of the $\mathbf{F}_{IB}$. It is noted that the velocity field $\mathbf{u}$ inside the sub-domain $\Omega_S$ should be theoretically uniform for a solid object, hence $\nabla \mathbf{u}$ and $\Delta \mathbf{u}$ at interior cells of $\Omega_S$ are zero according to Eqs (4)-(6). At the cells that cut across the object surface $\Gamma_{\Omega_S}$ there is a discretization residual as $\nabla \mathbf{u}$ and $\Delta \mathbf{u}$ at these cells are approximated using the normal velocity at cell surfaces locating outside the objects, which are in general not equal to the normal velocity at cell surfaces locating inside the objects. Numerically, this residual is expected to reduce with a finer mesh at the object surface. In contrast, the acceleration term, $(\mathbf{v}_0 - \mathbf{u})/dt$, at interior cells of $\Omega_S$ will generally be non-zero in the direction of movement. It is also noted that in the computation of $\mathbf{f}_{IB}$, the reconstructed velocity field of the reference mesh is used instead of the velocity of the object to maintain a consistency with the velocity subspaces. It is further shown in an numerical analysis in Section 3.3 below, the convection-diffusion is significantly smaller compared to the acceleration component of the $\mathbf{f}_{IB}$ and could be dropped out with insignificant effect to the PBROM results. This modification can significantly reduce computational cost in the on-line PBROM. This is because, in the on-line simulations, the sub-domain $\Omega_S$ is changing over time, hence $\mathbf{f}_{IB}$ needs to be re-evaluated at every time step. The computation of the convection-diffusion term in the $\mathbf{f}_{IB}$ requires reconstructions of full flow fields, computations of the divergence and Laplacian, as well as searching for computational cells inside $\Omega_S$. That increase the PBROM's runtime, hence degrade its efficiency. Although, this modification provides a simplification step to the PBROMs in certain cases, one can adopt full IB formula [28] for more accurate predictions of the force terms.

## 3 PBROM simulation results

In this section, the PBROM is tested for the predictions of transient flows past a single and two oscillating cylinders. The full CFD simulations for database generation are conducted in OpenFOAM. The numerical set-up for the two-dimensional flow past oscillating circular cylinders is sketched in Figure 2. The two cylinders have the same diameters of $D = 1\ m$. A constant velocity of $\mathbf{u}_{in} = (1,0)\ ms^{-1}$ is imposed at $\Gamma_{in}$. Zero gradient for velocity, $\nabla \mathbf{u} = 0$, is applied at $\Gamma_{top} \cup \Gamma_{bottom} \cup \Gamma_{out}$. On the cylinder surfaces, $\Gamma_{C_1}$ and $\Gamma_{C_2}$, oscillating velocities, i.e. $\mathbf{u}_1 = (u_{C_1}, v_{C_1})$ and $\mathbf{u}_2 = (u_{C_2}, v_{C_2})\ ms^{-1}$, are applied. Here the cylinders are allowed to move in the $y$-direction, i.e. $u_{C_1} = u_{C_2} = 0$ while $v_{C_1}$ and $v_{C_2}$ are computed from the prescribed cylinder positions $y_{C_1} = A_1 \sin(2\pi t/T_1)$ and $y_{C_2} = A_2 \sin(2\pi t/T_2)$. A zero gradient pressure condition, $\nabla p = 0$, is imposed at all boundaries, except at $\Gamma_{out}$ where $p = 0$ is imposed. The fluid density is $\rho = 1\ kgm^{-3}$. The kinematic viscosity is chosen as $\nu = 0.01\ m^2s^{-1}$, giving a Reynolds number of $Re = uD/\nu = 100$. A total of 124,338 computational cells are used to discretise the domain for the full CFD simulation.

The forces acting on the cylinder, $\mathbf{F}$, on the reference mesh comprise normal pressure force, $\mathbf{F}_p = \rho \sum_i \mathbf{s}_{f,i}(p_i - p_{ref})$ and tangential viscous force, $\mathbf{F}_v = \rho \sum_i \mathbf{s}_{f,i} \cdot (\nu_E \mathbf{R}_i)$, where $\mathbf{s}_{f,i}$ is the face area vector at and $\mathbf{R}_i$ is the deviatoric stress tensor of the velocity at the boundary face $i$ of the cylinder. These forces are computed in PBROM following the same OpenFOAM's function used in the CFD simulations. The $\mathbf{F}_{IB}$ forces are computed following Eq. (21).

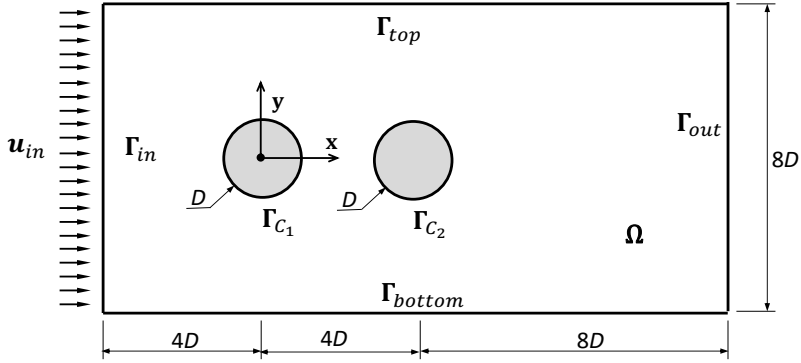

**Figure 2 Schematic description of the flow past oscillating cylinders problem.**

### 3.1 Flow past an oscillating cylinder at discrete amplitudes and frequencies

The PBROM with the reference mesh technique is first tested for a single oscillating cylinder. In this test case, the numerical set-up is sketched in Figure 2 with the second cylinder omitted. The PBROM model will be tested for two scenarios of discrete amplitudes and frequencies:

(1) The CFD simulation is carried out for 3 discrete oscillation amplitudes, $A_1 = 0.2, 0.5, 0.8\ m$, with single oscillation period $T_1 = 5\ s$ for the construction of the PBROM. The PBROM is then used to predict the flow field for: (Case 1a) oscillation amplitude coinciding with one used to build the ROM, $A_1 = 0.5\ m$; (Case 1b) oscillation amplitude within the range of those used to build the ROM, $A_1 = 0.35\ m$; and (Case 1c) oscillation amplitude outside the range of those used to build the ROM, $A_1 = 0.85\ m$. In all cases, the oscillation period remains at $T_1 = 5\ s$

(2) The CFD simulation is carried out for 3 discrete oscillation periods, $T_1 = 4.5, 5, 6\ s$, and a single oscillation amplitude $A_1 = 0.5\ m$ for the construction of the PBROM. The PBROM is then used to predict the flow field for: (Case 2a) oscillation period coinciding with one used to build the ROM, $T_1 = 4.5\ s$; (Case 2b) oscillation period within the range of those used to build the ROM, $T_1 = 5.2\ s$; and (Case 2c) oscillation period outside the range of those used to build the ROM, $T_1 = 6.2\ s$. In all cases, the oscillation amplitude remains at $A_1 = 0.5\ m$

A typical velocity field obtained from a full simulation is plotted with the deformed mesh in Figure 3. The PBROM model's initial conditions are derived from the CFD model's initial conditions by projecting them on the basis vectors. The "exact" solutions of model coefficients are the projections of CFD solutions on the basis vectors. Each PBROM simulation covers 20s of model time which is equivalent to 3-5 vortex shedding cycles.

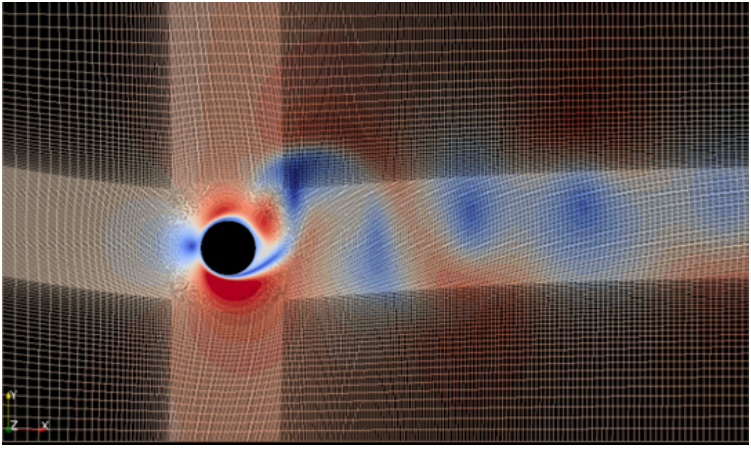

**Figure 3. Typical velocity field obtained from an OpenFOAM simulation of flow past an oscillating cylinder with deformed mesh. The prescribed cylinder position is $\mathbf{x}_{C_1} = (0, A_1 \sin(2\pi t/T_1))$**

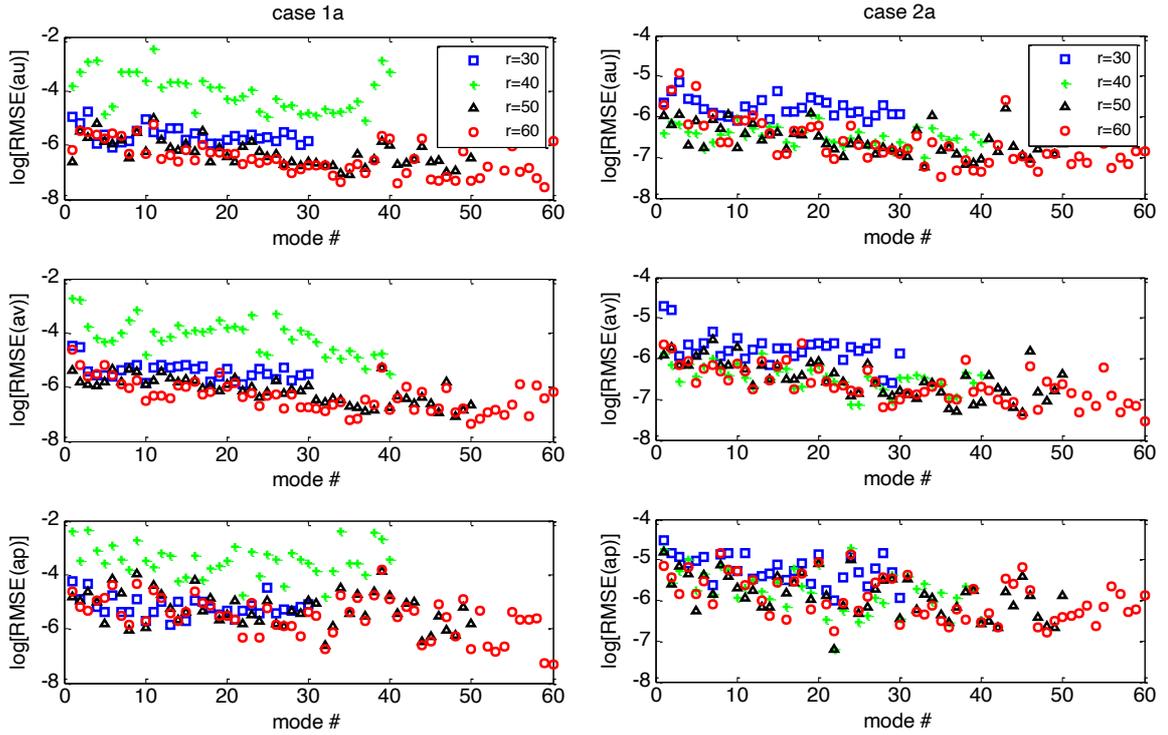

**Figure 4. RMSE for PBROM simulations of Case 1a and Case 2a.**

The number of POD modes used to construct the PBROM model are chosen based on a criterion that the accumulation of corresponding eigenvalues larger than 99.99% of the total eigenvalues, and based on the RMSE of modal coefficients with respect to numbers of modes. The RMSE is computed for every mode $a_i^\omega$ as $RMSE = \sqrt{1/n_T \sum_{j=1}^{n_T}[(a_{ij}^\omega)_P - (a_{ij}^\omega)_E]^2}$, where $(a_{ij}^\omega)_P$ and $(a_{ij}^\omega)_E$ are the predicted (PBROM) and exact (CFD) solutions of mode $a_i^\omega$ at time $t_j$ and $n_T$ is the number of data points. As shown in Figure 4, the RMSE of modal coefficients predicted by the 50- and 60-mode PBROM models are converged and are significantly lower than that of the 30- and 40-mode models. Therefore, 60 POD basis vectors are chosen.

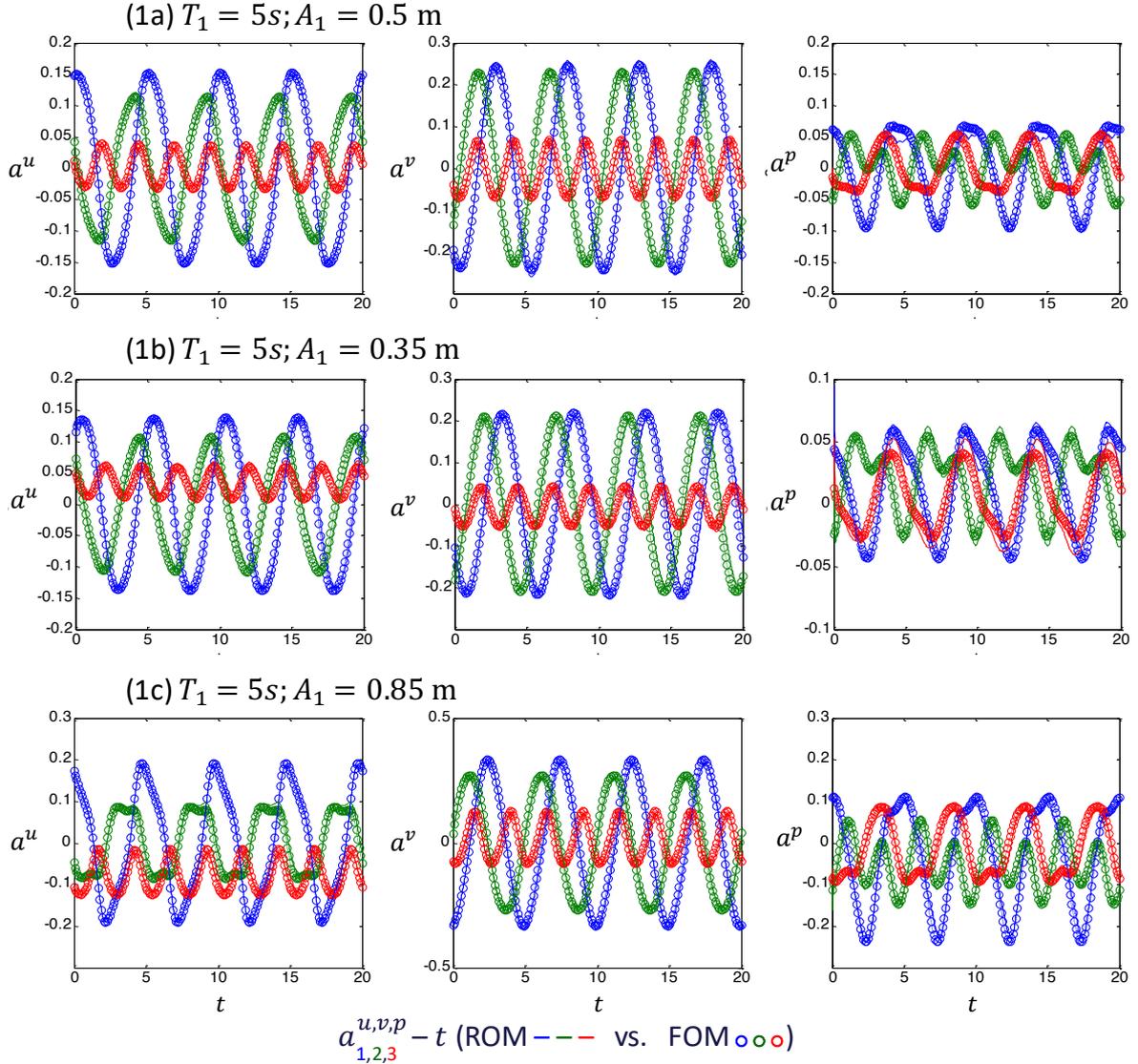

**Figure 5. Time-series of modal coefficients obtained from PBROM simulations (solid lines) and exact solutions (empty circle line) for Case 1a-c. Colour code: blue – mode 1, green – mode 2, red – mode 3.**

PBROM results of modal coefficients, forces on the cylinder surface for Case 1a-c and Case 2a-c and the respective CFD results are plotted in Figure 5 - Figure 7 for comparisons. Results indicate that PBROM models reproduce CFD solutions and predict unseen scenarios very well for both modal coefficients (Figure 5-6) over several vortex shedding cycles. The flow dynamics are dominated by the oscillation of the cylinder as one can observe from these figures. At the upper bounds of the amplitude and frequency ranges, the behaviours of mode 1 and 2 are quite different than that of the lower amplitude and frequency cases. The PBROM model performs very well in prediction for scenarios of new amplitude. Predictions at new frequencies are obviously more difficult. Mathematically, an interpolation (or extrapolation) between two frequencies would result in a primary and a secondary frequency. That reflects in the pressure coefficients in Case 2a and Case 2b. However, effects of the secondary frequency appear small in the velocity coefficients, which show not accumulating and not affecting the model stability and accuracy in overall.

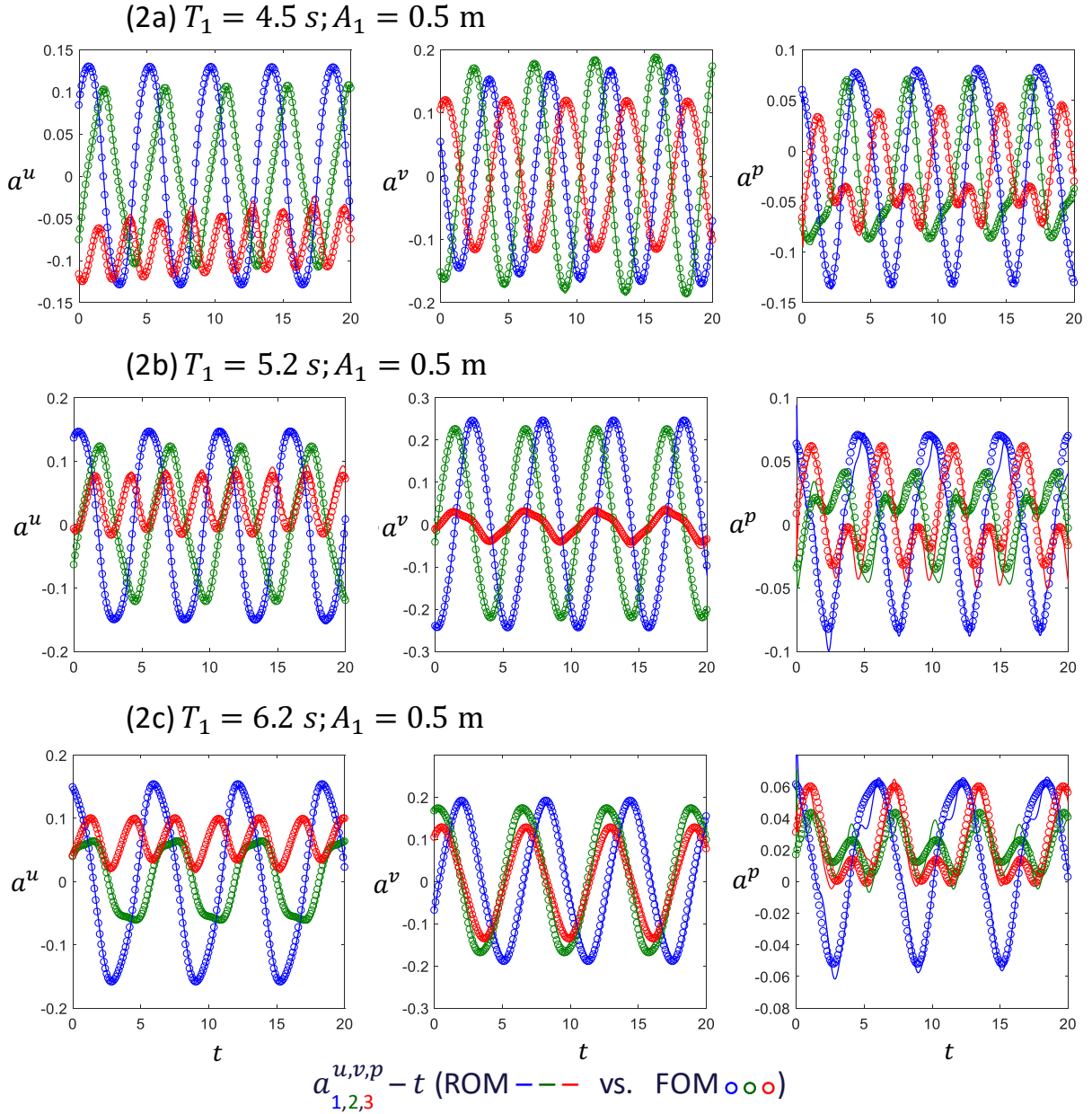

**Figure 6.** Time-series of modal coefficients obtained from PBROM simulations (solid lines) and exact solutions (empty circle line) for Case 2a-c. Colour code: blue – mode 1, green – mode 2, red – mode 3.

Figure 7 show that the PBROM models predict very well the forces acting on the cylinder. The predicted forces are as good as those for a stationary cylinder in [2] where the linear stochastic estimator (LSE) techniques was used to relate the pressure modal coefficients to the velocity modal coefficients to overcome the difficulty of developing a ROM for the Poisson's equation. Note that, in this PBROM model, the pressure modal coefficients are computed by solving the Poisson's equation with enforced boundary conditions on the moving cylinder together with the momentum equations.

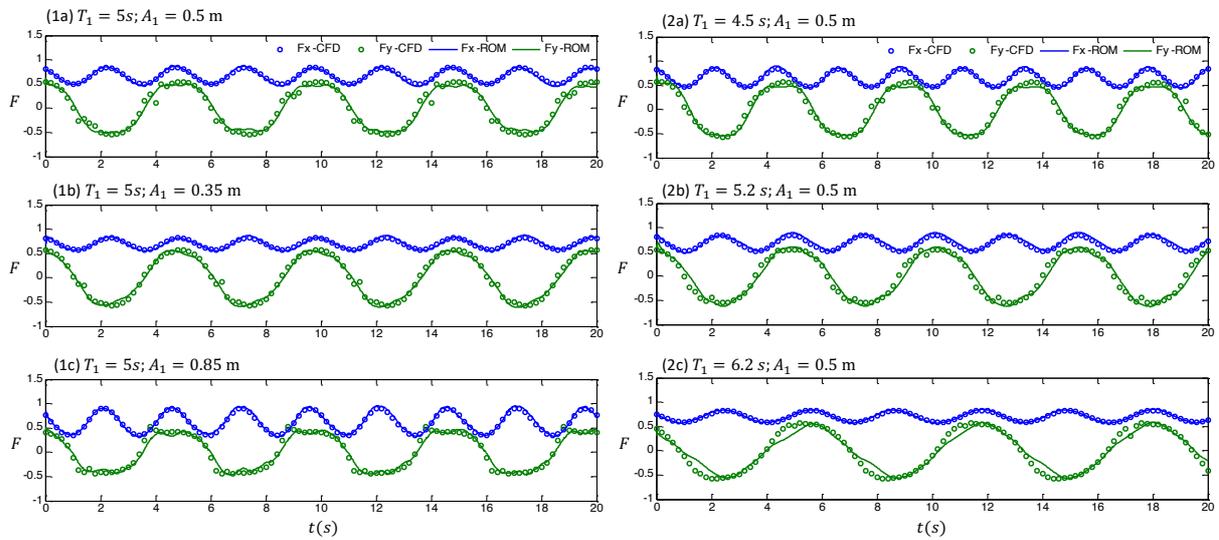

**Figure 7. Time-series of normalized forces on cylinder obatined from CFD and PBROM for Case 1a-c (left column) and Case 2a-c (right column).**

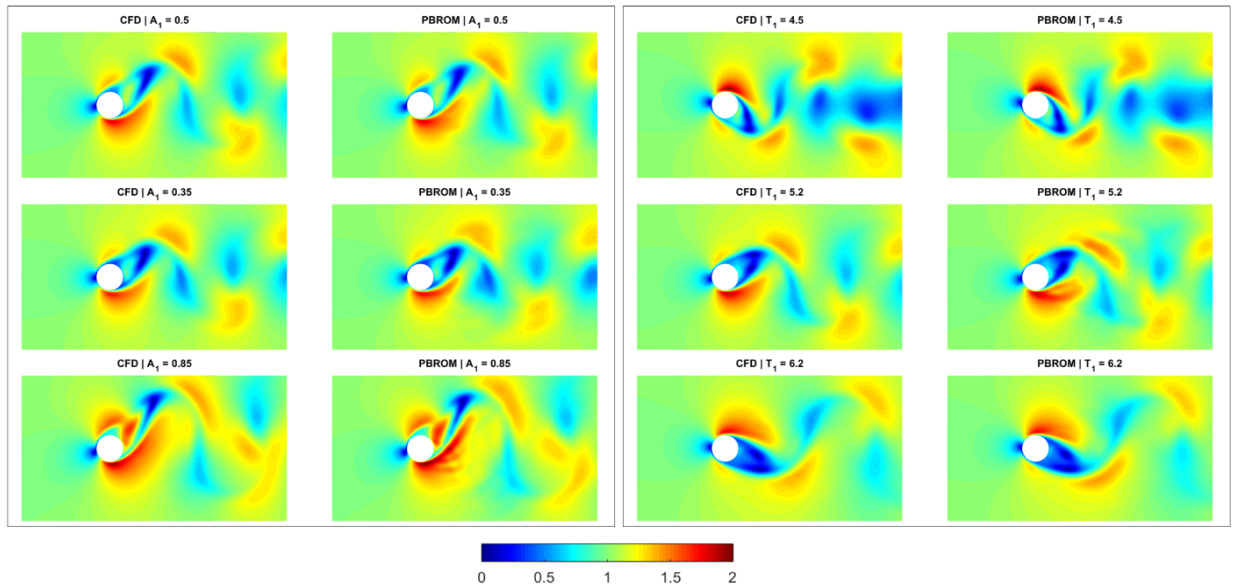

**Figure 8. Velocity magnitude (normalized) obtained from CFD and PBROM simulations at an arbitrary time for Case 1a-c (left box) and Case 2a-c (right box)**

Reconstructed flow fields from PBROM simulations for Case 1a-c and Case 2a-c are compared with respective CFD solutions at arbitrary times in Figure 8. The flow features evolving over time are captured accurately by the PBROM model for both seen and unseen scenarios. Predictions on the unseen scenarios that fall within the solution subspaces match well with CFD solutions while the extrapolations also achieve good solutions with only some minor differences.

The error appearing on the PBROM solution of Case 1c seems to come from the high frequency modes and not affecting the model stability.

### 3.2 Flow past an oscillating cylinder at varying amplitudes and frequencies

In the previous section, the PBROM models predict solutions at fixed single amplitudes or frequencies. In this section, the PBROM models with the ALE reference mesh technique are further tested for scenarios where amplitudes and frequencies vary during the simulations. The PBROM models are built at discrete points in amplitude and frequency spaces as before. This is more challenging as the PBROM models are required to capture the transient changes of flow dynamics over very short time periods when the oscillations of the cylinder change. Descriptions of Case 3 and Case 4 are presented below.

(3) The CFD simulation is carried out for 3 discrete oscillation amplitudes, $A_1 = 0.2, 0.5, 0.8\ m$, and a single period $T_1 = 5\ s$ for the construction of the PBROM. The PBROM is then used to predict the flow field for oscillation amplitude varies from 0.2 to $0.8\ m$ with increment of $0.2\ m$ after every 2 oscillation cycles (or $10\ s$). The period remains at $T_1 = 5\ s$.

(4) The CFD simulation is carried out for 3 discrete oscillation periods, $T_1 = 5, 5.5, 6\ s$, and a single amplitude $A_1 = 0.5\ m$ for the construction of the PBROM. The PBROM is then used to predict the flow field for oscillation period varies from 5 to $6\ s$ with increment of $0.25\ s$ after every 2 oscillation cycles. The amplitude remains at $A_1 = 0.5\ m$.

Figure 9 shows the positions of the cylinder of the two scenarios changing with time. In these two test cases, the amplitude and frequency of the cylinder are varied within the data ranges. Convergence tests suggest 40 modes for each variable to construct the reduced model. The PBROM simulation results and CFD solutions of modal coefficients are shown Figure 10. As one can observe, the PBROM models capture the dynamical evolution of major modes of velocity and pressure, including the spikes in pressure coefficients which corresponds to the jumps in the oscillation amplitude of the cylinder in Case 3. Underpredictions of pressure coefficients can be observed at some peaks of mode 1 and mode 3 curves in Case 3. In general, PBROM results match well with CFD solutions over long simulations.

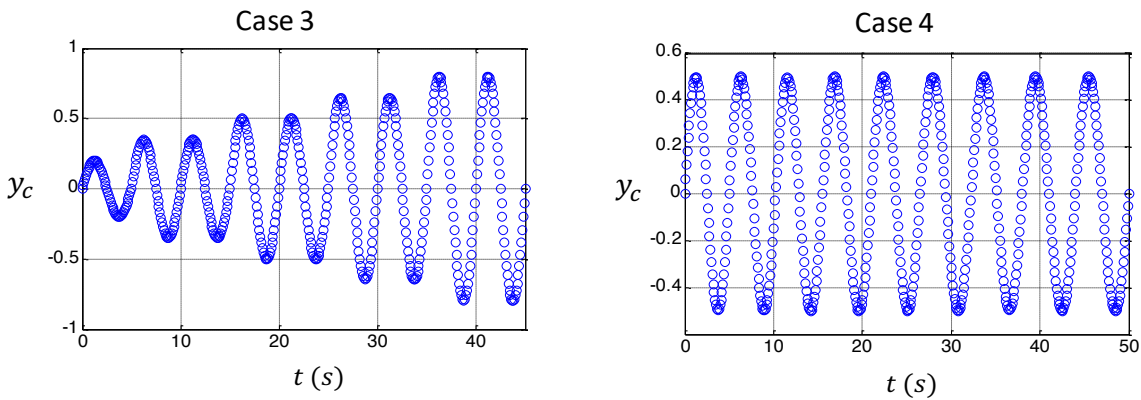

**Figure 9. Positions of the cylinder in the y-direction for the prediction Cases 3 (left) and Case 4 (right).**

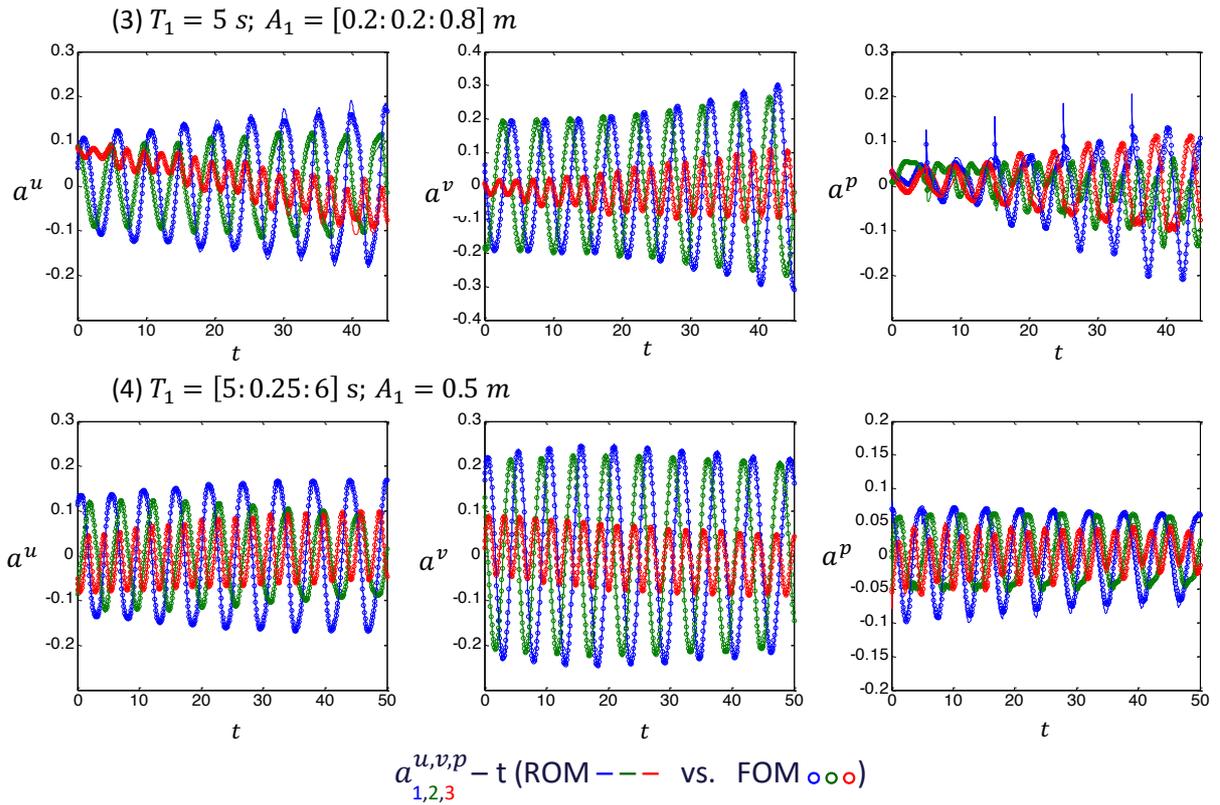

**Figure 10.** Time-series of modal coefficients for velocities and pressure obtained from PBROM simulations and exact solutions for Case 3-4. The empty circle symbols are "exact" solutions, solid lines are PBROM results. Colour represents mode number: blue – mode 1, green – mode 2, red – mode 3.

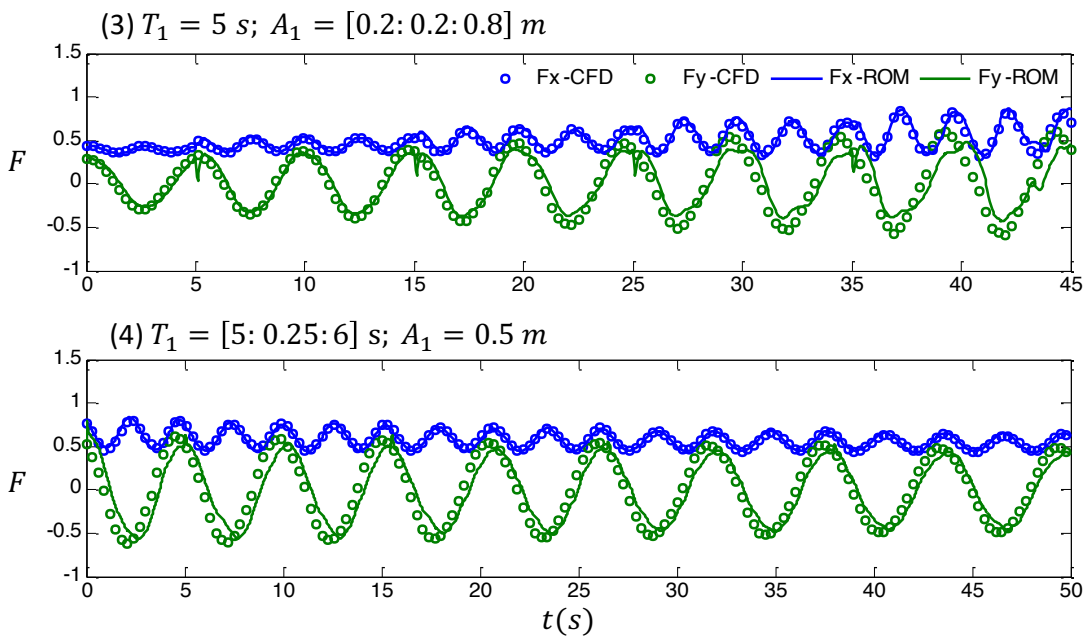

**Figure 11.** Time-series of normalized forces on the cylinder obatined from CFD and PBROM simulations solutions for Case 3 and Case 4.

The computed forces acting on the cylinder surface for Case 3 and Case 4 are compared with respective CFD solutions in Figure 11. The PBROM model of Case 3 predicts correctly the frequency and growing trends in both force components. The x-component matches well with the CFD solutions while the y-component is noticeably underpredicted when the oscillation amplitude increases. This underprediction of the y-component force are related to the underprediction of the pressure coefficients shown in Figure 10 for Case 3. Ghommem et al. [2] argued that the ROM model (19) only works well for low Reynolds steady flows and an extension of the ROM to a similar unsteady case is not straightforward due to the explicit nonlinear dependence of pressure field on velocity field. It must be noted that in the work by Ghommem et al., the pressure Poisson equation was not explicitly projected. In the author's previous work [32] it was showed that by using an explicit projection of the pressure Poisson equation with appropriate boundary conditions and the predictor-corrector scheme for solving the pressure equation, the PBROM can achieve good results for unsteady simulations. However, the PBROM model still takes more time to reach the quasi-steady states as compared to the full CFD counterparts. This explains for the under-growth of pressure coefficient and forces. Work to improve this issue is being carried out, including the adoption of an ML closure [14] to improve the model response. With no jump in the oscillation amplitude, the PBROM model of Case 4 predicts well both frequencies and amplitudes of the two forces components.

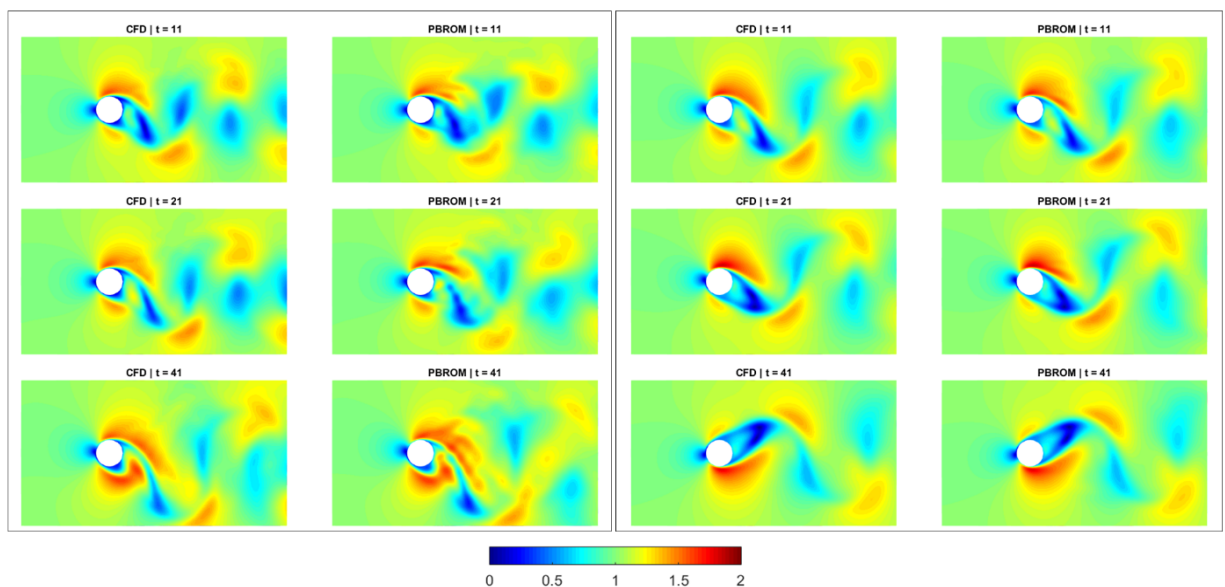

Figure 12. Velocity magnitude at different time instances obtained from CFD and PBROM simulations for Case 3 (left box) and Case 4 (right box)

Reconstructed flow fields from PBROM simulations at arbitrary time instances for Case 3 and Case 4 are compared with respective CFD solutions in Figure 12. The PBROM model for Case 4 is seen to produce very good predictions. The PBROM model for Case 3 captures all major features of the velocity fields, including magnitude, although there are some high order spatial fluctuations in velocity magnitude at the top left areas adjacent to the cylinder. These errors seem to come from the evolution of high-order modes. Correction methods such as the ML closure in [14] could also improve the predictions of high-order modes in velocity.

## 3.3 Verification of the IB-forcing method

As discussed in Section 2.4, the extension of the reference mesh technique to multiple moving objects is not straight forwards due to complex mesh deformations. The IB-like approach is, therefore, employed together with the reference mesh technique. However, the evaluation of the $\mathbf{f}_{IB}$ requires reconstructions of full flow fields, calls of differential operators and identifications of computational cells inside the moving objects every time step which will significantly increase the model runtime and degrade its efficiency. The below analysis is to demonstrate that dropping out the convection-diffusion component of the $\mathbf{f}_{IB}$ could significantly reduce model runtime while insignificantly affect the PBROM predictions numerically. In this test, a flow past a 2D oscillating cylinder with an amplitude of $A_1 = 0.2\ m$ is simulated. The problem is similar to the 2D setup in [21]. In this analysis, three different meshes are used to construct the PBROM models, namely M1, M2, M3. The number of modes ranges from 10 to 25. PBROM model without forcing, with acceleration term and with full forcing are tested.

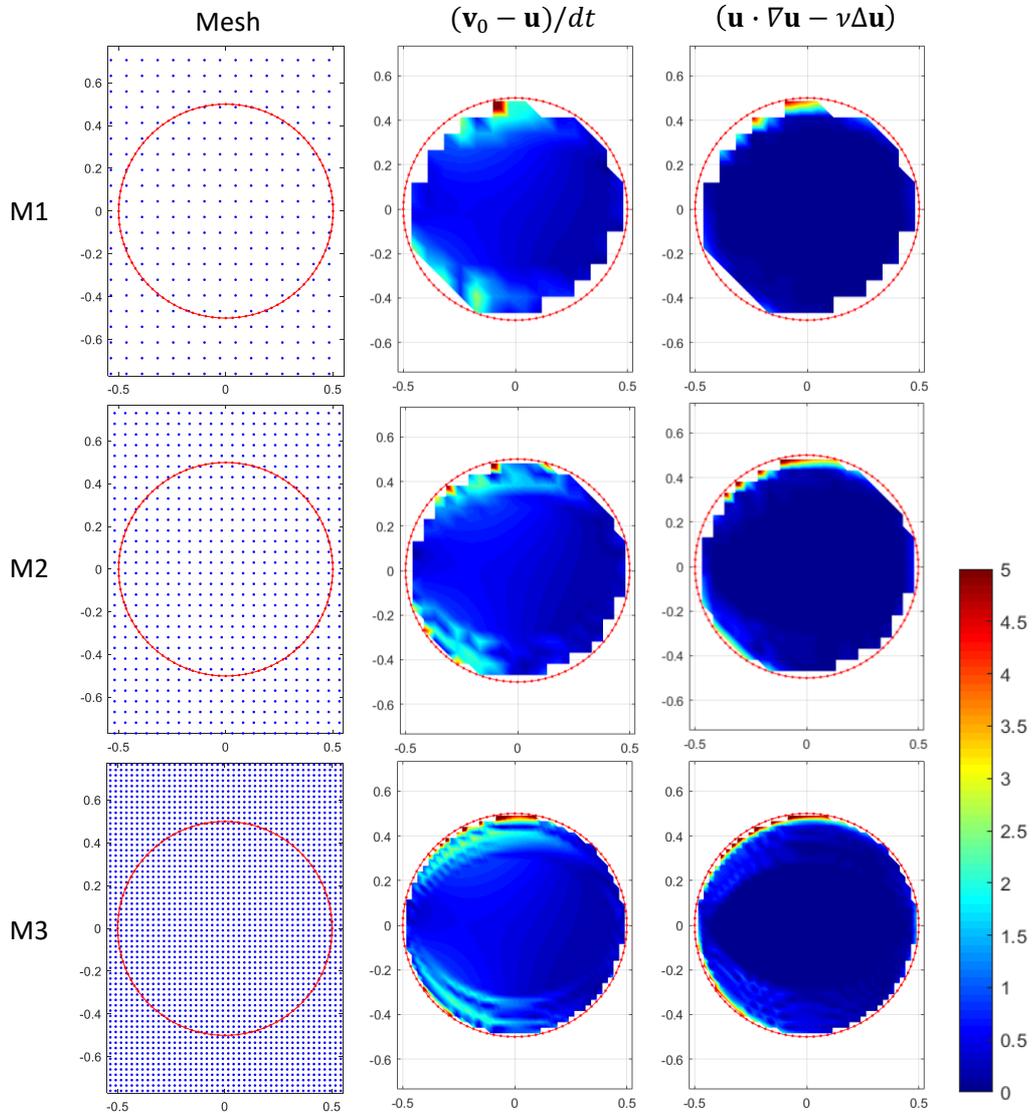

**Figure 13. Plots of mesh points ovelaid with object's surface, acceleration and convection-diffusion of F$_{IB}$ at computational cells inside the object (unit is ms$^{-2}$) for three different meshes M1, M2, M3.**

The computational meshes overlaid with object's surface are shown in Figure 13 with the fields of acceleration, $(\mathbf{v}_0 - \mathbf{u})/dt$, and the convection-diffusion, $(\mathbf{u} \cdot \nabla \mathbf{u} - \nu \Delta \mathbf{u})^n$, of $\mathbf{F}_{IB}$ at computational cells inside the object at an arbitrary time instance. On the three meshes, the number of computational cells locating inside the object are 149, 318, 1262 respectively and these numbers will vary when the object moves. In this test, no PBROM simulation is conducted, instead the $\mathbf{F}_{IB}$ is computed based on the velocity fields reconstructed from the truncated modes and the exact modal coefficients. One can see that the magnitudes of the convection-diffusion fields are significantly smaller than the acceleration fields and are generally close to zero. One can also observe that these fields are not uniform inside the object and larger magnitude patches appear nearer to the object surface. This is because the velocity fields reconstructed from truncated modes are not perfectly uniform and the difficulty of the POD method to approximate sharp velocity changes at the object surface. The non-uniformity of the velocity field reflects directly in the fields of acceleration (as $\mathbf{v}_0$ is uniform and $dt$ is constant). The non-zero values of the convection-diffusion term are not only the result of the POD error in approximating the velocity but also of the discretization residual of $\nabla \mathbf{u}$ and $\Delta \mathbf{u}$ at surface-crossing cells (as discussed in Section 2.4). The error is seen to be higher but only localized near the object surface. Those issues will be addressed in the future.

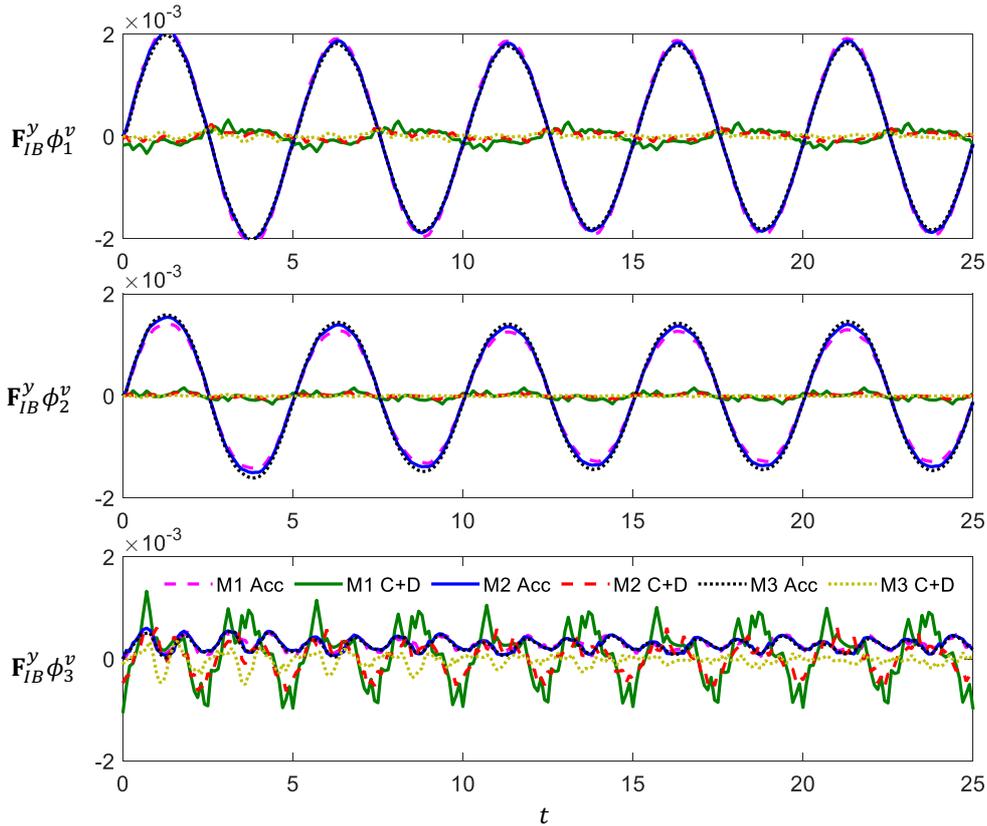

**Figure 14.** Projections of the acceleration (Acc) and the convection-diffusion (C+D) of $\mathbf{F}_{IB}$ on the first three modes of the vertical velocity components. Results are shown for three different meshes M1, M2, M3

The projections of the y-component of $(\mathbf{v}_0 - \mathbf{u})/dt$ and $(\mathbf{u} \cdot \nabla \mathbf{u} - \nu \Delta \mathbf{u})^t$ on the first three modes of the vertical velocity components, $F_Y \phi^v_{1,2,3}$, are analysed (as the object moves in the y-direction only). Shown in Figure 14, the convection-diffusion components in the projections on the first two modes are approximately one order smaller than the acceleration components. The projections on mode 3 of both components are at the same order but both are generally small. While the acceleration terms are generally consistent over three mesh resolutions, the convection-diffusion term sees significant changes: the magnitudes are smaller for finer meshes and the changes are more significant in mode 3. The errors in the convection-diffusion component are observed in the projection of higher order modes, but are generally small and reflecting the non-uniform velocity in the interior of the object and the numerical error encountered at surface-crossing cells which tend to be picked up in the high order modes.

The model runtimes of the PBROM with different of $\mathbf{F}_{IB}$ models on different meshes are shown in Table 1. Generally, runtime increase with the increase of the number of modes. There is no significant difference in runtimes of the model with no forcing over the three meshes. Adding the acceleration forcing term, the runtimes increase significantly and the differences over the meshes are also significant. The M3 mesh shows ~10 s increases of runtime (~55–90 %) as compared to M2 while M2 shows smaller increases in runtime as compared to M1 (~3–20 %). One can also observe significant increases (~120–320 s or ~520–1450 %) in runtime of the PBROM model with full $\mathbf{F}_{IB}$ on mesh M3 as compared to the model with acceleration forcing only. On mesh M2, the runtime increase is less but still significant (~32–62 s or ~270–375 %). On mesh M1, the runtime increase is ~180 %. The conclusion is that the computation of the convection-diffusion term takes a significant portion of the total runtime and increases with larger problems (having more modes or on a finer mesh). Hence dropping it out will help to improve the performance of the $\mathbf{F}_{IB}$ model. Nevertheless, with the full $\mathbf{F}_{IB}$ model, i.e. Model C, the PBROM simulation is still much faster compared to the CFD simulation, which is approximately 3 hours on a 24-cores workstation (or 72 CPU hours). This motivates developments of full $\mathbf{F}_{IB}$ model into PBROM in the future.

Table 1. Runtime (seconds) of the PBROM with different of $\mathbf{F}_{IB}$ models on different meshes.

| # modes | (A) No forcing | | | (B) with Acceleration $F_{IB}$ | | | (C) with full $F_{IB}$ | | |
|---|---|---|---|---|---|---|---|---|---|
| | M1 | M2 | M3 | M1 | M2 | M3 | M1 | M2 | M3 |
| 10 | 2.5 | 2.6 | 2.7 | 10.1 | 12.1 | 22.7 | 28.8 | 44.4 | 140.8 |
| 15 | 3.1 | 3.1 | 3.2 | 12.2 | 12.6 | 23.8 | 33.3 | 53.2 | 196.9 |
| 20 | 3.6 | 3.6 | 3.8 | 13.2 | 13.9 | 24.1 | 37.6 | 63.8 | 288 |
| 25 | 4 | 4.1 | 4.3 | 15.3 | 16.5 | 25.6 | 42.5 | 78.3 | 397.7 |

The PBROM models constructed without and with the two IB forcing models are further tested in actual simulations. The PBROM models are built on mesh M2 with 25 modes. Results in Figure 15 show that the $\mathbf{F}_{IB}$ model with only acceleration term (B) is as good as the full model with both acceleration and convection-diffusion terms (C). Both PBROM models B and C perform well and results are very close to the CFD. Model A shows quick decay of the modal coefficients, obviously due to the lack of forcing from the moving object. The phase portraits (or limit cycles) of $a_1^v - a_2^v$ are compared to the CFD result in Figure 16. With either forcing models employed, the PBROM captures accurately the dynamics response between the first two major modes.

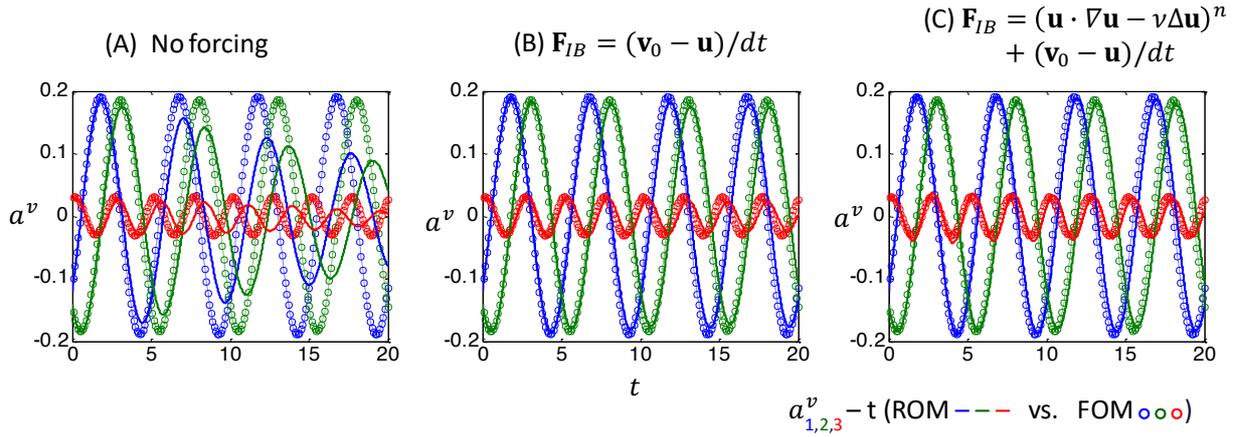

Figure 15. Time-series of modal coefficients for velocities and pressure obtained from PBROM simulations (on mesh M2) and exact solutions of flow past a single oscillating cylinder, $A_1 = 0.2\ m$. The empty circle symbols are "exact" solutions, solid lines are PBROM results. Colour represents mode number: blue – mode 1, green – mode 2, red – mode 3.

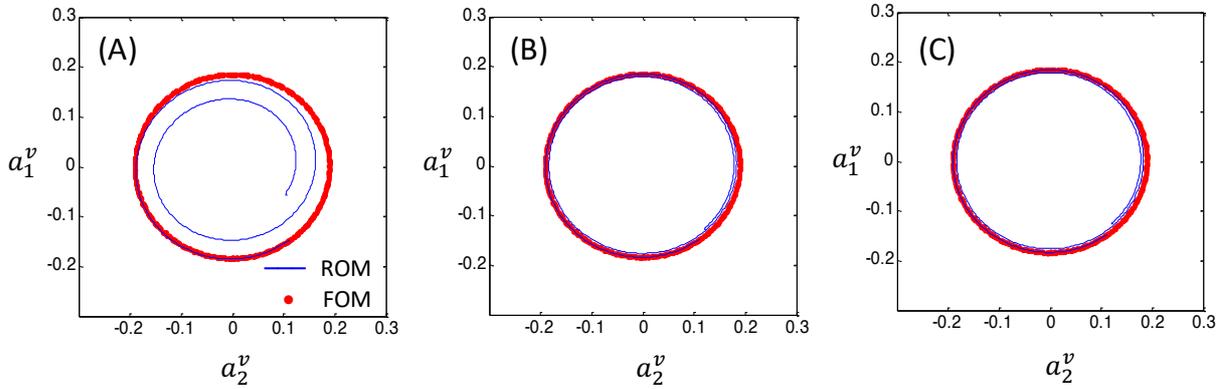

Figure 16. The phase portraits of $a_1^v - a_2^v$ computed from the three PBROM models (mesh M2) and from full order CFD simulation (FOM).

### 3.4 Flow past two oscillating cylinders

The PBROM model with reference mesh and IB-like approaches is demonstrated for Case 5: flow past two oscillating cylinders. The oscillation amplitudes and periods of the two cylinders are $A_1 = 0.3\ m$, $T_1 = 5\ s$ and $A_2 = 0.2\ m$, $T_2 = 5.5\ s$. A convergence test similar to one presented in the above section suggests 60 modes for each variable for the construction of the reduced model. The PBROM simulations are performed with the same oscillation amplitudes and periods of the two cylinders. The mesh resolution is the same as that of the M2 mesh.

Figure 17 shows the reference mesh fixed to the 1st cylinder and zoomed in at the area of the 2nd cylinder together with the cylinder positions at different time instances. One can observe that, at different cylinder positions, the numbers of interior cells are different. This will have certain impacts to the accuracy of the $\mathbf{f}_{IB}$ in the PBROM model. The maximum displacement of the cylinder centre is $0.5\ m$ in relative to the mesh. The runtimes of the full order CFD and the PBROM simulations for 45s of model time are approximately 126 hrs and 60 s respectively.

Results of the PBROM models are presented in Figure 18 - Figure 20 and compared with CFD results.

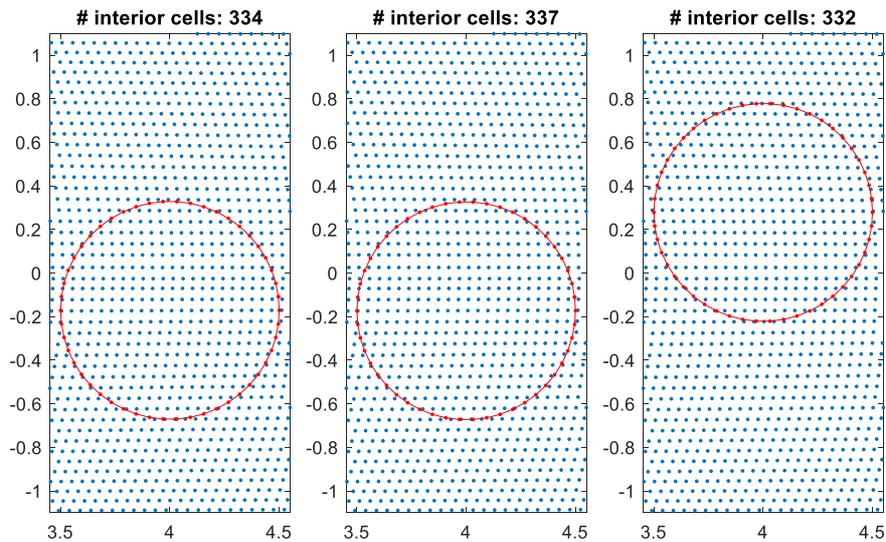

Figure 17. Plots of mesh points ovelaid with object's surface at different time instances.

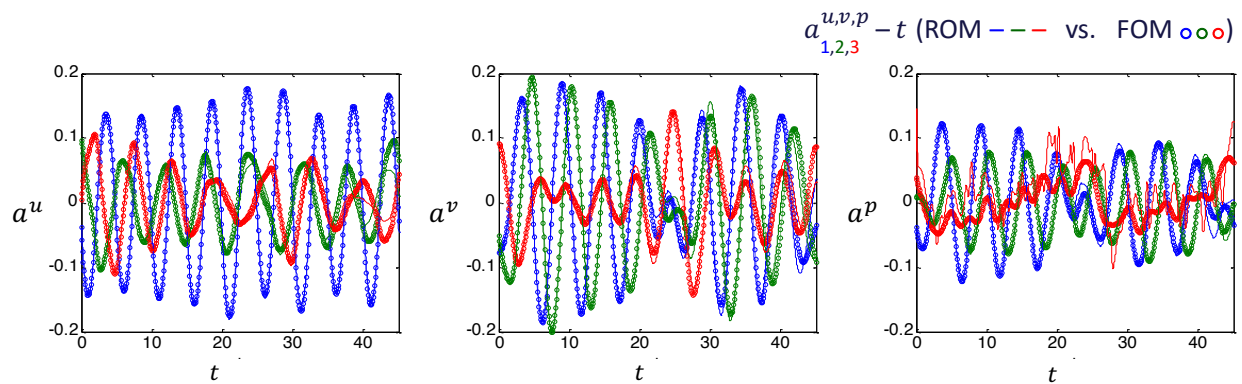

Figure 18. Time-series of modal coefficients for velocities and pressure obtained from PBROM simulations and CFD for Case 5. The empty circle symbols are from CFD, solid lines are PBROM results. Colour represents mode number: blue – mode 1, green – mode 2, red – mode 3.

Simulation results of modal coefficients for velocities and pressure are shown in Figure 18. The PBROM models capture well behaviours of velocity modes. The response the first two pressure modes is also captured well. The 3rd mode of pressure generally follows the response in the CFD result although there are some fluctuations which seems not affecting the accuracy of first two pressure modes and the velocity modes.

Forces acting on the two cylinders are shown in Figure 19. The forces acting on the 1st cylinder are captured very accurately by the PBROM models. This is expected as the reference mesh is attached to this cylinder and the result of PBROM model is accurate as demonstrated in the single cylinder test cases. The forces acting on the 2nd cylinder are more complex than those on the 1st cylinder due to the wake of the 1st cylinder. The PBROM model with $\mathbf{f}_{IB}$ forcing appears to follow very well the dynamical behaviours of the forces on the 2nd cylinder, although some slight overshoots/undershoots are observed in y-component.

Reconstructed flow fields from PBROM simulations at arbitrary time instances are compared with respective CFD solutions in Figure 20. The flow fields with major features are predicted well by the PBROM models. Flow field in the domain that covers the 2nd cylinder are shown in Figure 21. The velocity inside the cylinder is fairly uniform and the sharp jumps of velocity across the object surface are also captured reasonably well.

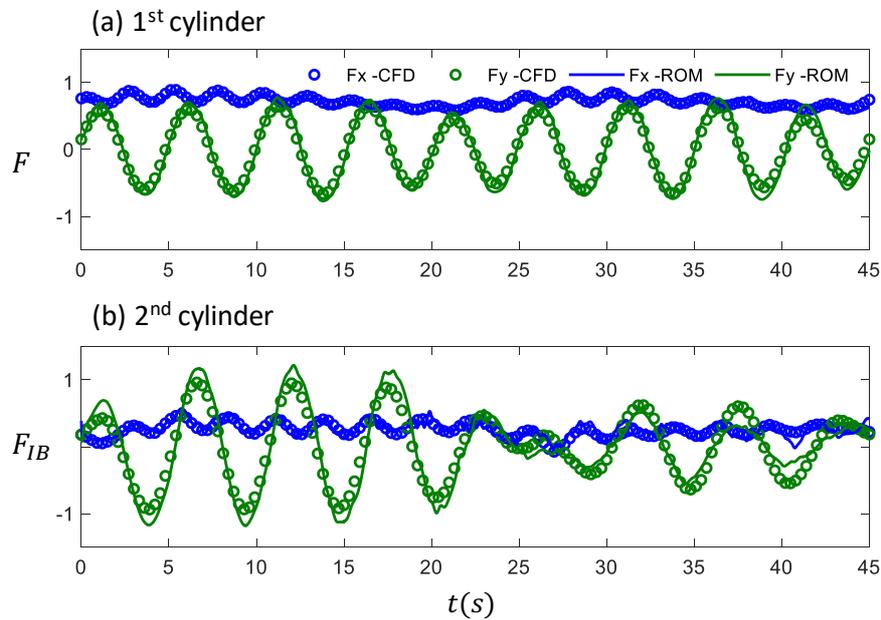

**Figure 19. Time-series of forces (normalized) on the cylinder obtained from CFD and PBROM simulations solutions for Case 5.**

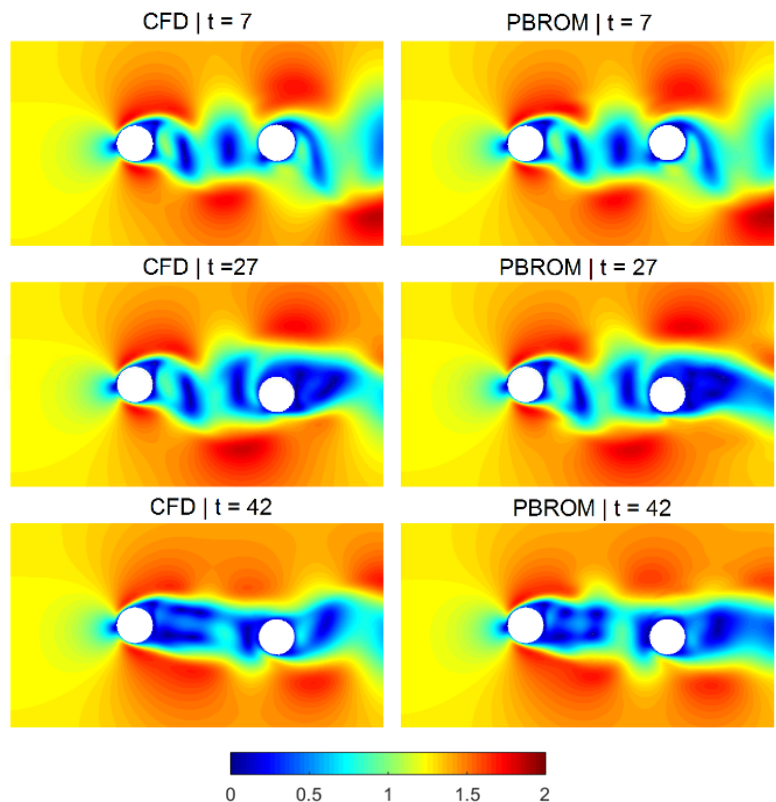

**Figure 20. Velocity magnitudes at different time instances from CFD and PBROM simulations for Case 5**

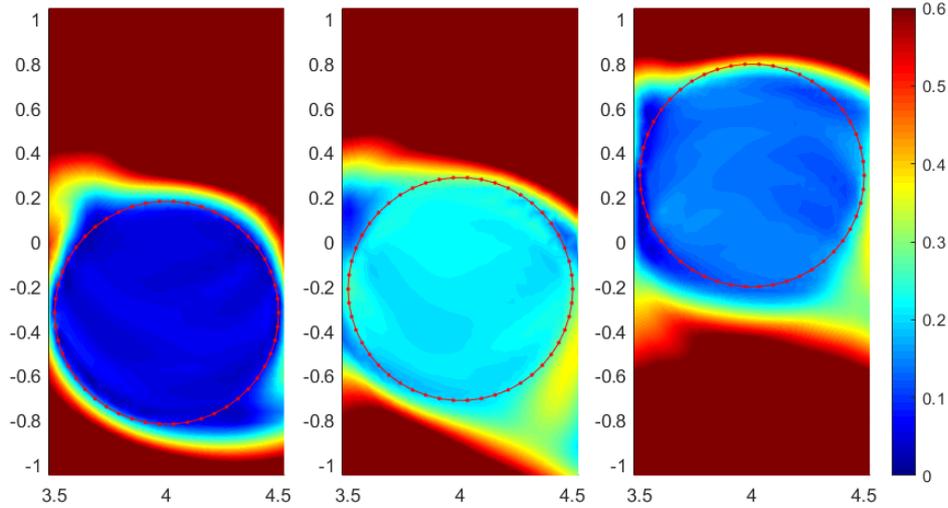

Figure 21. Velocity magnitude at different time instances from PBROM zoomed in at the 2nd cylinder

# 4 Conclusions

In this paper, a projection-based reduced order modelling (PBROM) approach for transient simulations of multiple moving objects in nonlinear crossflows is presented. A combination of moving reference mesh and simplified IB techniques are used to model complex moving boundaries in the reduced model. The implementation of the moving reference mesh technique is straightforward but requires a capability of accurately imposing arbitrary time-dependent boundary conditions at the moving boundaries in the PBROM model. The present PBROM model implemented in OpenFoam and the moving reference mesh technique is shown accurate for the moving object where a non-deforming mesh can be attached to it. The model can predict accurately the transient flow dynamics and forces for unseen scenarios. The simplified IB technique was shown working well in combination with the moving mesh technique for a complex scenario involving multiple moving objects. The implementation of the IB method is simple but effective. Simulation time could be reduced by more than 180% on a coarse mesh as compared to an existing method and could be more than 1000% on a fine mesh. Although the PBROM simulations are much faster compared to the CFD simulations while capture well flow fields and forces, there are rooms for further improvements to the IB method. These include using a more comprehensive IB methods to calculate the forcing terms, special treatments at boundary-crossing cells or simply use a finer mesh for the PBROM with a mindfulness of computational cost that may incur. These implementations will be considered in a follow-up development. These methods will also be tested for more practical three-dimensional objects, induced motions or higher Re flows.

**Appendix**

Details of functions $f_{\mathbf{u}}, g_{\mathbf{u}}, f_p, g_p$ of the ODEs (19)-(20) are given as follows

$$a_t^u = (A_0^u + C_{L0}^u - C_{P0}^u) + (A_1^u + C_{L1}^u)a^u + A_2^u a^v + A_3^u a^w + a^{u^T} B_1^u a^u + a^{v^T} B_2^u a^u + \\ a^{w^T} B_3^u a^u - C_{P1}^u a^p + G_L^u b^u + b^{u^T} G_1^u b^u + b^{v^T} G_2^u b^u + b^{w^T} G_3^u b^u - G_P^u b^p \tag{24}$$

$$a_t^v = (A_0^v + C_{L0}^v - C_{P0}^v) + A_1^v a^u + (A_2^v + C_{L1}^v)a^v + A_3^v a^w + a^{u^T} B_1^v a^v + a^{v^T} B_2^v a^v + \\ a^{w^T} B_3^v a^v - C_{P1}^v a^p + G_L^v b^v + b^{u^T} G_1^v b^v + b^{v^T} G_2^v b^v + b^{w^T} G_3^v b^v - G_P^v b^p \tag{25}$$

$$a_t^w = (A_0^w + C_{L0}^w - C_{P0}^w) + A_1^w a^u + A_2^w a^v + (A_3^w + C_{L1}^w)a^w + a^{u^T} B_1^w a^w + a^{v^T} B_2^w a^w + \\ a^{w^T} B_3^w a^w - C_{P1}^w a^p + G_L^w b^w + b^{u^T} G_1^w b^w + b^{v^T} G_2^w b^w + b^{w^T} G_3^w b^w - G_P^w b^w \tag{26}$$

$$E_1 a^p = \left(-E_0 + \frac{D_{D0}}{dt} + D_{L0} + D_0\right) + \left(\frac{D_{D1}}{dt} + D_{L1} + D_1\right)a^u + \left(\frac{D_{D2}}{dt} + D_{L2} + D_2\right)a^v + \left(\frac{D_{D3}}{dt} + D_{L3} + D_3\right)a^w + a^{u^T} D_{11} a^u + a^{v^T} D_{22} a^v + a^{w^T} D_{33} a^w + a^{v^T} D_{21} a^u + a^{w^T} D_{32} a^v + \\ a^{u^T} D_{13} a^w + \left(\frac{H_{D1}}{dt} + H_{L1}\right)b^u + \left(\frac{H_{D2}}{dt} + H_{L2}\right)b^v + \left(\frac{H_{D3}}{dt} + H_{L3}\right)b^w + b^{u^T} H_{11} b^u + \\ b^{v^T} H_{22} b^v + b^{w^T} H_{33} b^w + b^{v^T} H_{21} b^u + b^{w^T} H_{32} b^v + b^{u^T} D_{13} b^w - H_L b^p \tag{27}$$

where

$[A_0^u]_i = -(\mathbf{u}_0 \cdot \nabla u_0, \phi_i^u)$, $[A_1^u]_{ij} = -(\phi_j^u \cdot \nabla_x u_0, \phi_i^u) - (\mathbf{u}_0 \cdot \nabla \phi_j^u, \phi_i^u)$,
$[A_2^u]_{ij} = -(\phi_j^v \cdot \nabla_y u_0, \phi_i^u)$, $[A_3^u]_{ij} = -(\phi_j^w \cdot \nabla_z u_0, \phi_i^u)$,
$[B_1^u]_{ijk} = -(\phi_j^u \cdot \nabla_x \phi_k^u, \phi_i^u)$, $[B_2^u]_{ijk} = -(\phi_j^v \cdot \nabla_y \phi_k^u, \phi_i^u)$, $[B_3^u]_{ijk} = -(\phi_j^w \cdot \nabla_z \phi_k^u, \phi_i^u)$,
$[C_{L0}^u]_i = (\nu \Delta u_0, \phi_i^u)$, $[C_{L1}^u]_{ij} = (\nu \Delta \phi_j^u, \phi_i^u)$,
$[C_{P0}^u]_i = (\nabla_x p_0, \phi_i^u)$, $[C_{P1}^u]_{ij} = (\nabla_x \phi_j^p, \phi_i^u)$,
$[G_1^u]_{ijk} = -(g_j^u \cdot \nabla_x g_k^u, \phi_i^u)$, $[G_2^u]_{ijk} = -(g_j^v \cdot \nabla_y g_k^u, \phi_i^u)$, $[G_3^u]_{ijk} = -(g_j^w \cdot \nabla_z g_k^u, \phi_i^u)$,
$[G_{L1}^u]_{ij} = (\nu \Delta g_j^u, \phi_i^u)$, $[G_{P1}^u]_{ij} = (\nabla_x g_j^p, \phi_i^u)$,

$[A_0^v]_i = -(\mathbf{u}_0 \cdot \nabla v_0, \phi_i^v)$, $[A_1^v]_{ij} = -(\phi_j^u \cdot \nabla_x v_0, \phi_i^v)$,
$[A_2^v]_{ij} = -(\phi_j^v \cdot \nabla_y v_0, \phi_i^v) - (\mathbf{u}_0 \cdot \nabla \phi_j^v, \phi_i^v)$, $[A_3^v]_{ij} = -(\phi_j^w \cdot \nabla_z v_0, \phi_i^v)$,
$[B_1^v]_{ijk} = -(\phi_j^u \cdot \nabla_x \phi_k^v, \phi_i^v)$, $[B_2^v]_{ijk} = -(\phi_j^v \cdot \nabla_y \phi_k^v, \phi_i^v)$, $[B_3^v]_{ijk} = -(\phi_j^w \cdot \nabla_z \phi_k^v, \phi_i^v)$,
$[C_{L0}^v]_i = (\nu \Delta v_0, \phi_i^v)$, $[C_{L1}^v]_{ij} = (\nu \Delta \phi_j^v, \phi_i^v)$,
$[C_{P0}^v]_i = (\nabla_y p_0, \phi_i^v)$, $[C_{P1}^v]_{ij} = (\nabla_y \phi_j^p, \phi_i^v)$,
$[G_1^v]_{ijk} = -(g_j^u \cdot \nabla_x g_k^v, \phi_i^v)$, $[G_2^v]_{ijk} = -(g_j^v \cdot \nabla_y g_k^v, \phi_i^v)$, $[G_3^v]_{ijk} = -(g_j^w \cdot \nabla_z g_k^v, \phi_i^v)$,
$[G_{L1}^v]_{ij} = (\nu \Delta g_j^v, \phi_i^v)$, $[G_{P1}^v]_{ij} = (\nabla_y g_j^p, \phi_i^v)$,

$[A_0^w]_i = -(\mathbf{u}_0 \cdot \nabla w_0, \phi_i^w)$, $[A_1^w]_{ij} = -(\phi_j^w \cdot \nabla_x w_0, \phi_i^w)$,
$[A_2^w]_{ij} = -(\phi_j^v \cdot \nabla_y w_0, \phi_i^w)$, $[A_3^w]_{ij} = -(\phi_j^w \cdot \nabla_z w_0, \phi_i^w) - (\mathbf{u}_0 \cdot \nabla \phi_j^w, \phi_i^w)$,
$[B_1^w]_{ijk} = -(\phi_j^u \cdot \nabla_x \phi_k^w, \phi_i^w)$, $[B_2^w]_{ijk} = -(\phi_j^v \cdot \nabla_y \phi_k^w, \phi_i^w)$, $[B_3^w]_{ijk} = -(\phi_j^w \cdot \nabla_z \phi_k^w, \phi_i^w)$,
$[C_{L0}^w]_i = (\nu \Delta w_0, \phi_i^w)$, $[C_{L1}^w]_{ij} = (\nu \Delta \phi_j^w, \phi_i^w)$,
$[C_{P0}^w]_i = (\nabla_z p_0, \phi_i^w)$, $[C_{P1}^w]_{ij} = (\nabla_z \phi_j^p, \phi_i^w)$,
$[G_1^w]_{ijk} = -(g_j^u \cdot \nabla_x g_k^w, \phi_i^w)$, $[G_2^w]_{ijk} = -(g_j^v \cdot \nabla_y g_k^w, \phi_i^w)$, $[G_3^w]_{ijk} = -(g_j^w \cdot \nabla_z g_k^w, \phi_i^w)$,
$[G_{L1}^w]_{ij} = (\nu \Delta g_j^w, \phi_i^w)$, $[G_{P1}^w]_{ij} = (\nabla_z g_j^p, \phi_i^w)$,

$[E_0]_i = \left(\frac{1}{\rho} \Delta p_0, \phi_i^p\right)$, $[E_1]_{ij} = \left(\frac{1}{\rho} \Delta \phi_j^p, \phi_i^p\right)$, $[H_L]_{ij} = \left(\frac{1}{\rho} \Delta g_j^p, \phi_i^p\right)$,
$[D_{D0}]_i = (\nabla \cdot \mathbf{u}_0, \phi_i^p)$, $[D_{D1}]_{ij} = (\nabla_x \phi_j^u, \phi_i^p)$, $[D_{D2}]_{ij} = (\nabla_y \phi_j^v, \phi_i^p)$, $[D_{D3}]_{ij} = (\nabla_z \phi_j^w, \phi_i^p)$,
$[D_{L0}]_i = (\nu \Delta (\nabla \cdot \mathbf{u}_0), \phi_i^p)$,
$[D_{L1}]_{ij} = (\nu \Delta (\nabla_x \phi_j^u), \phi_i^p)$, $[D_{L2}]_{ij} = (\nu \Delta (\nabla_y \phi_j^v), \phi_i^p)$, $[D_{L3}]_{ij} = (\nu \Delta (\nabla_z \phi_j^w), \phi_i^p)$,
$[D_0]_i = -(\nabla_x (\mathbf{u}_0 \cdot \nabla u_0) + \nabla_y (\mathbf{u}_0 \cdot \nabla v_0) + \nabla_z (\mathbf{u}_0 \cdot \nabla w_0), \phi_i^p)$,
$[D_1]_{ij} = -(\nabla_x (\phi_j^u \cdot \nabla u_0) + \nabla_y (\phi_j^u \cdot \nabla v_0) + \nabla_z (\phi_j^u \cdot \nabla w_0) + \nabla_x (\mathbf{u}_0 \cdot \nabla \phi_j^u), \phi_i^p)$,
$[D_2]_{ij} = -(\nabla_x (\phi_j^v \cdot \nabla_y u_0) + \nabla_y (\phi_j^v \cdot \nabla_y v_0) + \nabla_z (\phi_j^v \cdot \nabla_y w_0) + \nabla_y (\mathbf{u}_0 \cdot \nabla \phi_j^v), \phi_i^p)$,

$[\boldsymbol{D}_3]_{ij} = -\left(\nabla_x(\phi_j^w \cdot \nabla_z u_0) + \nabla_y(\phi_j^w \cdot \nabla_z v_0) + \nabla_z(\phi_j^w \cdot \nabla_z w_0) + \nabla_z(\mathbf{u}_0 \cdot \nabla \phi_j^w), \phi_i^p\right),$

$[\boldsymbol{D}_{11}]_{ijk} = -\left(\nabla_x(\phi_j^u \cdot \nabla_x \phi_k^u), \phi_i^p\right), \quad [\boldsymbol{D}_{22}]_{ijk} = -\left(\nabla_y(\phi_j^v \cdot \nabla_y \phi_k^v), \phi_i^p\right),$

$[\boldsymbol{D}_{33}]_{ijk} = -\left(\nabla_z(\phi_j^w \cdot \nabla_z \phi_k^w), \phi_i^p\right),$

$[\boldsymbol{D}_{21}]_{ijk} = -\left(\nabla_x(\phi_j^u \cdot \nabla_y \phi_k^v) + \nabla_y(\phi_j^v \cdot \nabla_x \phi_k^u), \phi_i^p\right),$

$[\boldsymbol{D}_{32}]_{ijk} = -\left(\nabla_y(\phi_j^v \cdot \nabla_z \phi_k^w) + \nabla_z(\phi_j^w \cdot \nabla_y \phi_k^v), \phi_i^p\right),$

$[\boldsymbol{D}_{13}]_{ijk} = -\left(\nabla_z(\phi_j^w \cdot \nabla_x \phi_k^u) + \nabla_x(\phi_j^u \cdot \nabla_z \phi_k^w), \phi_i^p\right),$

$[\boldsymbol{H}_{D1}]_{ij} = \left(\nabla_x g_j^u, \phi_i^p\right), \quad [\boldsymbol{H}_{D2}]_{ij} = \left(\nabla_y g_j^v, \phi_i^p\right), \quad [\boldsymbol{H}_{D3}]_{ij} = \left(\nabla_z g_j^w, \phi_i^p\right),$

$[\boldsymbol{H}_{L1}]_{ij} = \left(\nu\Delta(\nabla_x g_j^u), \phi_i^p\right), \quad [\boldsymbol{H}_{L2}]_{ij} = \left(\nu\Delta(\nabla_y g_j^v), \phi_i^p\right), \quad [\boldsymbol{H}_{L3}]_{ij} = \left(\nu\Delta(\nabla_z g_j^w), \phi_i^p\right),$

$[\boldsymbol{H}_{11}]_{ijk} = -\left(\nabla_x(g_j^u \cdot \nabla_x g_k^u), \phi_i^p\right), \quad [\boldsymbol{H}_{22}]_{ijk} = -\left(\nabla_y(g_j^v \cdot \nabla_y g_k^v), \phi_i^p\right),$

$[\boldsymbol{H}_{33}]_{ijk} = -\left(\nabla_z(g_j^w \cdot \nabla_z g_k^w), \phi_i^p\right),$

$[\boldsymbol{H}_{21}]_{ijk} = -\left(\nabla_x(g_j^u \cdot \nabla_y g_k^v) + \nabla_y(g_j^v \cdot \nabla_x g_k^u), \phi_i^p\right),$

$[\boldsymbol{H}_{32}]_{ijk} = -\left(\nabla_y(g_j^v \cdot \nabla_z g_k^w) + \nabla_z(g_j^w \cdot \nabla_y g_k^v), \phi_i^p\right),$

$[\boldsymbol{H}_{13}]_{ijk} = -\left(\nabla_z(g_j^w \cdot \nabla_x g_k^u) + \nabla_x(g_j^u \cdot \nabla_z g_k^w), \phi_i^p\right),$

Here, $\nabla_x, \nabla_y, \nabla_z$ are the gradient components in each coordinate, i.e. $\nabla = (\nabla_x, \nabla_y, \nabla_z)$.